\documentclass{sig-alternate-10pt}
\usepackage[T1]{fontenc}
\usepackage{times}  
\usepackage{epsfig}
\usepackage{afterpage}
\usepackage{tabularx}
\usepackage{graphicx}
\usepackage{balance}
\usepackage{color}
\usepackage{xspace}
\usepackage{thumbpdf}
\usepackage{listings}
\usepackage{verbatim}
\usepackage{color}
\usepackage[hidelinks]{hyperref}
\definecolor{darkred}{rgb}{0.7,0,0}
\definecolor{darkgreen}{rgb}{0,0.5,0}
\hypersetup{colorlinks=true,
        linkcolor=darkred,
        citecolor=darkgreen}
\usepackage{booktabs}
\usepackage{colortbl}
\usepackage[inline]{aplcomments}
\usepackage{inconsolata}
\usepackage{paralist}
\usepackage{xspace}
\usepackage{listings}
\usepackage{siunitx}
\usepackage{breakurl}
\usepackage{inconsolata}
\usepackage{longtable}
\usepackage{placeins}
\usepackage{caption, subcaption}
\usepackage{pbox}
\usepackage{pifont}
\usepackage{tablefootnote}
\newcommenter{ak}{1.0,1.0,0.3}
\newcommenter{ac}{0.4,1.0,1.0}

\lstdefinestyle{customc}{
 belowcaptionskip=1\baselineskip,
 breaklines=true,
 language=C,
 escapeinside={@}{@},
 showstringspaces=false,
 basicstyle=\small\ttfamily,
 keywordstyle=\bfseries\color{green!40!black},
 commentstyle=\itshape\color{purple!40!black},
 stringstyle=\color{orange},
 directivestyle=\color{brown},
}

\lstdefinestyle{customctable}{
 aboveskip=-\medskipamount,
 belowskip=-\medskipamount,
 language=C,
 escapeinside={@}{@},
 showstringspaces=false,
 basicstyle=\scriptsize\ttfamily,
 keywordstyle=\bfseries\color{green!40!black},
 commentstyle=\itshape\color{purple!40!black},
 stringstyle=\color{orange},
 directivestyle=\color{brown},
}

\def\compactify{\itemsep=0pt \topsep=0pt \partopsep=0pt \parsep=0pt}
\let\latexusecounter=\usecounter

\newenvironment{CompactEnumerate}
  {\def\usecounter{\compactify\latexusecounter}
   \begin{enumerate}}
  {\end{enumerate}\let\usecounter=\latexusecounter}

  \usepackage{hyperref}
  
\sloppypar
\begin{document}

\date{}

\title{Programmable Packet Scheduling}
\author{
\alignauthor \fontsize{10.7}{9.9}\selectfont Anirudh Sivaraman\textsuperscript{*}, Suvinay Subramanian\textsuperscript{*}, Anurag Agrawal\textsuperscript{\dag}, Sharad Chole\textsuperscript{\ddag}, Shang-Tse Chuang\textsuperscript{\ddag}, Tom Edsall\textsuperscript{\ddag}, Mohammad Alizadeh\textsuperscript{*}, Sachin Katti\textsuperscript{+}, Nick McKeown\textsuperscript{+}, Hari Balakrishnan\textsuperscript{*}\\
\affaddr{\fontsize{10.7}{9.9}\selectfont \textsuperscript{*}MIT CSAIL, \textsuperscript{\dag}Barefoot Networks, \textsuperscript{\ddag}Cisco Systems, \textsuperscript{+}Stanford University}\\
}

\maketitle


\begin{abstract}

Switches today provide a small set of scheduling algorithms. While we can tweak
scheduling parameters, we cannot modify algorithmic logic, or add a
completely new algorithm, after the switch has been designed. This paper
presents a design for a {\em programmable} packet scheduler, which allows
scheduling algorithms---potentially algorithms that are unknown today---to be
programmed into a switch without requiring hardware redesign.

Our design builds on the observation that scheduling algorithms make two
decisions: {\em in what order} to schedule packets and {\em when} to schedule
them. Further, in many scheduling algorithms these decisions can be made when
packets are enqueued. We leverage this observation to build a programmable
scheduler using a single abstraction: the push-in first-out queue (PIFO), a
priority queue that maintains the scheduling order and time for such algorithms.

We show that a programmable scheduler using PIFOs lets us program a wide
variety of scheduling algorithms. We present a detailed hardware design for
this scheduler for a 64-port 10 Gbit/s shared-memory switch with <4\% chip area
overhead on a 16-nm standard-cell library.  Our design lets us  program many
sophisticated algorithms, such as a 5-level hierarchical scheduler
with programmable scheduling algorithms at each level.

\end{abstract}

\section{Introduction}
\label{s:intro}

Today's line-rate switches provide a menu of scheduling algorithms: typically,
a combination of Deficit Round Robin~\cite{drr}, strict priority scheduling,
and traffic shaping. A network operator can configure parameters in these
algorithms. However, an operator cannot change the core algorithmic logic in an existing
scheduling algorithm, or program a new one, without building new switch
hardware.

By contrast, with a {\em programmable} packet scheduler, network operators
would be able to deploy custom scheduling algorithms to better meet application
requirements, e.g., minimizing flow completion times~\cite{pFabric} using
Shortest Remaining Processing Time~\cite{srpt}, flexible bandwidth allocation
across flows or tenants~\cite{faircloud, eyeq} using Weighted Fair
Queueing~\cite{wfq}, or minimizing tail packet delays~\cite{intserv} using
Least Slack Time First~\cite{lstf}. With a programmable packet scheduler,
switch designers would implement scheduling algorithms as programs atop a
programmable substrate. Moving scheduling algorithms into software makes it
much easier to build and verify algorithms in comparison to implementing the
same algorithms as rigid hardware IP.

This paper presents a design for programmable packet scheduling in line-rate
switches. Our design is motivated by the observation that all scheduling
algorithms make two key decisions: first, in what order should packets be
scheduled, and second, at what time should each packet be scheduled.
Furthermore, in many scheduling algorithms, these two decisions can be made
when a packet is enqueued. This observation was first made in a recent position
paper~\cite{pifo_hotnets}. The same paper also proposed the {\em push-in first-out
queue (PIFO)}~\cite{pifo} abstraction for maintaining the scheduling order or
scheduling time for packets, when these can be determined on enqueue. A PIFO
is a priority queue data structure that allows elements to be pushed into an
arbitrary location based on an element's {\em rank}, but always dequeues
elements from the head.

Building on the PIFO abstraction, this paper presents the detailed design,
implementation, and analysis of feasibility of a programmable packet scheduler.
To program a PIFO, we develop the notion of a {\em scheduling transaction}---a
small program to compute an element's rank in a PIFO. We present a rich
programming model built using PIFOs and scheduling transactions
(\S\ref{s:pifo}) and show how to program a diverse set of scheduling algorithms
in the model (\S\ref{s:expressive}): Weighted Fair Queueing~\cite{wfq}, Token
Bucket Filtering~\cite{tbf}, Hierarchical Packet Fair Queueing~\cite{hpfq},
Class-Based Queueing~\cite{cbq, cbq_impl}, Least-Slack Time-First~\cite{lstf},
Stop-and-Go Queueing~\cite{stopngo}, the Rate-Controlled Service
Disciplines~\cite{rcsd}, and fine-grained priority scheduling (e.g., Shortest
Job First, Shortest Remaining Processing Time, Least Attained Service, and
Earliest Deadline First).

Until now, all line-rate implementations of these scheduling algorithms---if
they exist at all---have been hard-wired into switch hardware. We also describe the
limits of the PIFO abstraction (\S\ref{ss:limitations}) by presenting examples
of scheduling algorithms that can't be programmed using a PIFO.

We present a detailed hardware design for a programmable scheduler using PIFOs
(\S\ref{s:design}).  We have implemented this design and synthesized it to an
industry-standard 16 nm standard-cell library (\S\ref{s:hardware}). We find,
contrary to conventional wisdom~\cite{drr, sfq}, that transistor technology has
scaled to a point where the sorting operation at the core of a PIFO is
surprisingly cheap. As a consequence, we show that it is feasible to build a
programmable scheduler, which
\begin{CompactEnumerate}
  \item supports 5-level hierarchical scheduling, where the scheduling
    algorithms at each level are programmable.
  \item runs at a clock frequency of 1 GHz---sufficient for a 64-port
    shared-memory switch with a 10 Gbit/s line rate per port.
  \item incurs <4\% chip area overhead relative to a shared-memory switch
    supporting a small set of scheduling algorithms.
  \item handles the same buffering requirements as a typical shared-memory switch
    today~\cite{trident} (about 60K packets and 1K flows).
\end{CompactEnumerate}

\section{A programming model for packet scheduling}
\label{s:pifo}

This section introduces the basic abstractions and programming model
we use to express packet scheduling algorithms. The key idea
underlying this programming model is that any scheduling algorithm
makes two decisions: the {\em order} in which packets are
scheduled, and the {\em time} at which they are scheduled. These two
decisions capture work-conserving and non-work-conserving scheduling
algorithms respectively. Further, in many practical scheduling
algorithms, the order and time can be determined when a packet is
enqueued into the packet buffer.

Our programming model is built around this intuition and has two
components:
\begin{CompactEnumerate}
\item The {\em push-in first-out queue (PIFO)}~\cite{pifo} data structure that
  maintains the scheduling order or scheduling time for algorithms where these
  can be determined at enqueue. A PIFO is a priority queue that allows elements
  to be pushed into an arbitrary location on enqueue based on an element's {\em
  rank}, but dequeues elements from the head.\footnote{We use the term rank
  instead of priority to avoid confusion with strict priority scheduling.
  Throughout this paper, lower ranks are dequeued first from the PIFO.} A PIFO
  breaks ties between elements with the same rank in the order in which they were
  enqueued.
\item A set of operations on the PIFO data structure called {\em
    transactions} that compute an element's rank before enqueuing it
    into a PIFO.
\end{CompactEnumerate}

We now describe the three main abstractions of our programming
model. First, we show how to use a {\em scheduling transaction} to
program simple work-conserving scheduling algorithms using a single
PIFO~(\S\ref{ss:wfq}).  Second, we generalize to a {\em tree} of
scheduling transactions to program hierarchical work-conserving
scheduling algorithms~(\S\ref{ss:hpfq}). Third, we augment nodes of
this tree with a {\em shaping transaction} to program
non-work-conserving scheduling algorithms~(\S\ref{ss:hshaping}).

\subsection{Scheduling transactions}
\label{ss:wfq}

A scheduling transaction is a block of code that is executed for each packet
before enqueueing it into a PIFO. The scheduling transaction computes a rank for
the packet. This rank then determines the position in the PIFO where the packet
is enqueued. Scheduling transactions are an instance of packet
transactions~\cite{domino_arxiv} --- blocks of code that are atomic and
isolated from other such transactions. Packet transactions guarantee that any
visible state is equivalent to a serial execution of these transactions across
consecutive packets.

Scheduling transactions can be used to program work-conserving
scheduling algorithms. In particular, a single scheduling transaction
(and PIFO) is sufficient to program any scheduling algorithm where
the relative scheduling order of packets already in the buffer does
not change with the arrival of future packets.

Take Weighted Fair Queueing (WFQ)~\cite{wfq} as an example. WFQ
provides weighted max-min allocation of link capacity across
flows\footnote{We use the term `flow' to generically describe a set
  packets with a common attribute. For example, a flow could be
  packets destined to the same subnet, or video packets, or a TCP
  connection.} sharing a link. Practical approximations to WFQ
include Deficit Round Robin (DRR)~\cite{drr}, Stochastic Fairness
Queueing (SFQ)~\cite{sfq}, and Start-Time Fair Queueing
(STFQ)~\cite{stfq}. We consider STFQ here.

Before a packet is enqueued, STFQ computes a {\em virtual start time} for that
packet as the maximum of the {\em virtual finish time} of the previous packet in that
packet's flow and the current value of the {\em virtual time} (a single state
variable that tracks the virtual start time of the last dequeued packet).
Packets are then scheduled in order of increasing virtual start times.  To
program STFQ using a PIFO, we use the scheduling transaction shown in
Figure~\ref{fig:sched_trans}.

Across all transactions in the paper, we use the notation {\tt p.x} to
refer to a packet field and set {\tt p.rank} to the desired value at
the end of the transaction based on the computations in the
transaction.

\begin{figure}
\begin{lstlisting}[style=customc]
f = flow(p) // compute flow from packet p
if f in last_finish
  p.start = max(virtual_time, last_finish[f])
else
  p.start = virtual_time
last_finish[f] = p.start + p.length / f.weight
p.rank = p.start
\end{lstlisting}
\caption{Scheduling transaction for STFQ}
\label{fig:sched_trans}
\end{figure}

\subsection{Tree of scheduling transactions}
\label{ss:hpfq}

\begin{figure}
\centering
\includegraphics[width=0.5\textwidth]{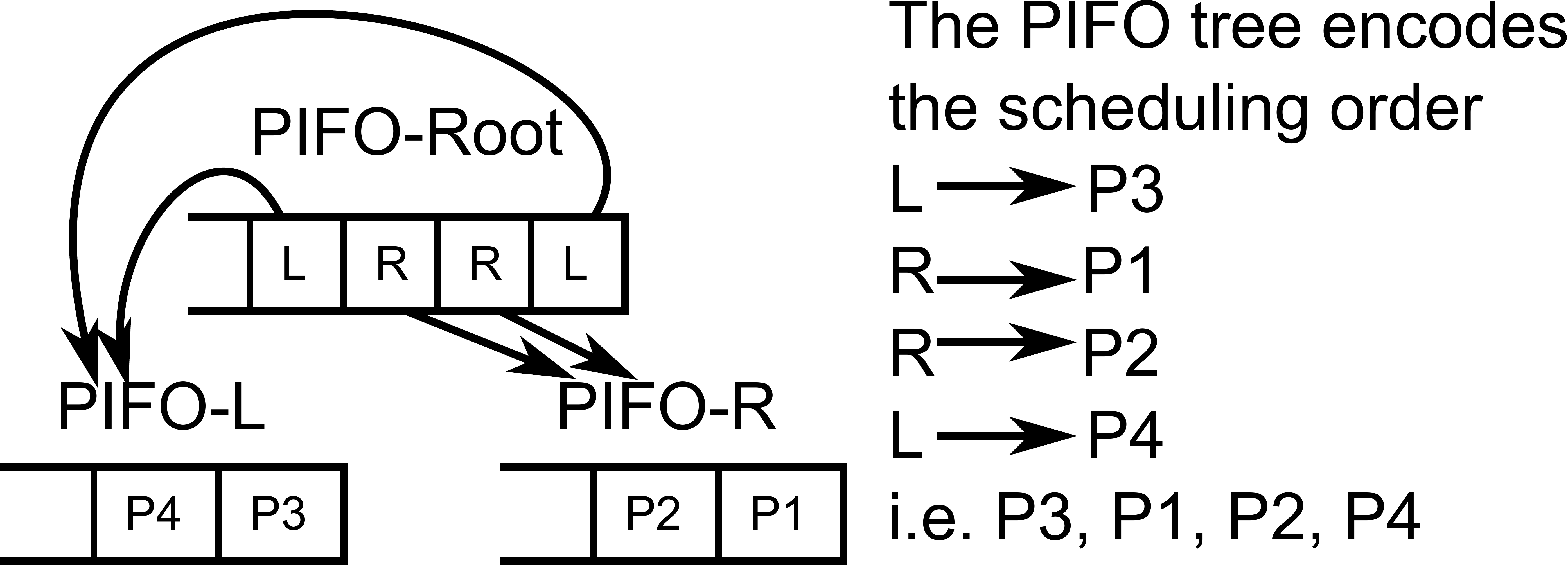}
\caption{PIFO trees encode instantaneous scheduling order.}
\label{fig:pifo_encoding}
\end{figure}

\begin{figure*}
\begin{subfigure}[b]{.2\textwidth}
\includegraphics[width=\textwidth]{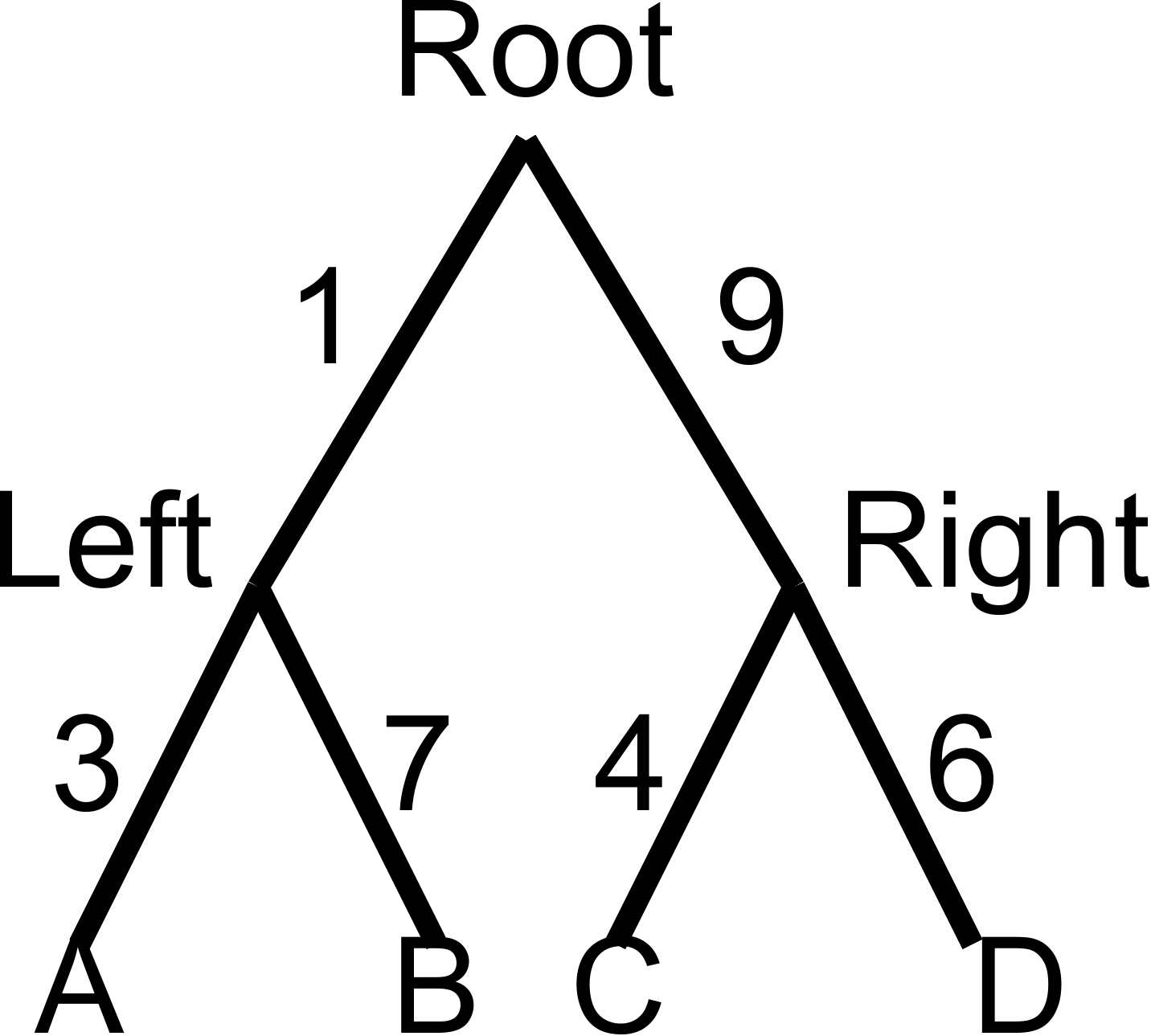}
\caption{Algorithm}
\label{fig:hpfq_algo}
\end{subfigure}
\vrule
\begin{subfigure}[b]{.3\textwidth}
\includegraphics[width=\textwidth]{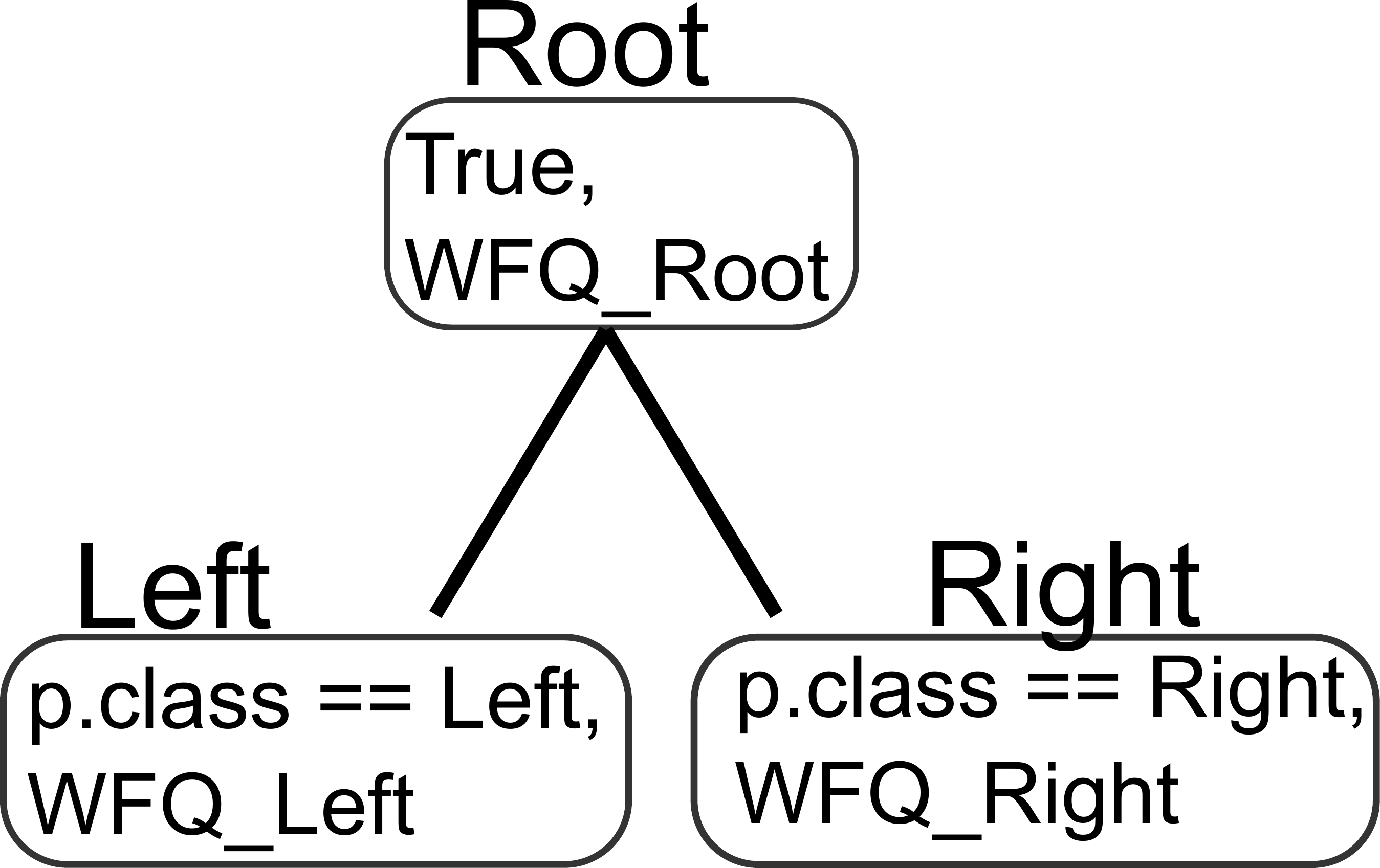}
\caption{Tree of PIFOs}
\label{fig:hpfq_tree}
\end{subfigure}
\vrule
\begin{subfigure}[b]{.5\textwidth}
\begin{center}
\begin{lstlisting}[style=customc]
f = flow(p)
// compute flow from packet p:
// (Left / Right for WFQ_Root)
// (A / B for WFQ_Left)
// (C / D for WFQ_Right)
if f in last_finish
  p.start = max(virtual_time, last_finish[f])
else
  p.start = virtual_time
last_finish[f] = p.start + p.length / f.weight
p.rank = p.start
\end{lstlisting}
\end{center}
\caption{Scheduling transaction for WFQ\_Root, WFQ\_Left, and
  WFQ\_Right.}
\label{fig:hpfq_trans}
\end{subfigure}
\caption{Programming HPFQ using PIFOs}
\label{fig:hpfq}
\end{figure*}

Scheduling algorithms that require changing the relative scheduling
order of already buffered packets with the arrival of new packets cannot be
implemented with a single scheduling transaction and PIFO. An
important class of such algorithms are {\em hierarchical} schedulers
that compose multiple scheduling policies at different levels of
hierarchy. We introduce the idea of a {\em tree of scheduling
  transactions} to program such algorithms.

To illustrate this idea, consider Hierarchical Packet Fair Queueing
(HPFQ)~\cite{hpfq}. HPFQ first
apportions link capacity between classes, then recursively between sub classes
belonging to each class, all the way down to the leaf nodes.
Figure~\ref{fig:hpfq_algo} provides an example scheduling hierarchy, the
numbers on the edges indicating the relative weights of child nodes with
respect to their parent's fair scheduler.  HPFQ cannot be realized using a
single scheduling transaction and PIFO because the relative scheduling order
of packets that are already buffered can change with future packet
arrivals (see Section 2.2 of the HPFQ paper~\cite{hpfq} for an example).

HPFQ {\em can}, however, be realized using a tree of PIFOs, with a
scheduling transaction attached to each PIFO in the tree. To see how, observe that HPFQ
simply executes some variant of WFQ at each level of the hierarchy,
with each node using WFQ to pick among its children. As we showed in
\S\ref{ss:wfq}, a single PIFO encodes the instantaneous
scheduling order for WFQ, i.e. the scheduling order if there are no
further arrivals. Similarly, a tree of PIFOs
(Figure~\ref{fig:pifo_encoding}), where each PIFO's elements are
either packets or references to other PIFOs can be used to encode the
instantaneous scheduling order of HPFQ (and other hierarchical
scheduling algorithms) as follows. First, inspect the root PIFO to
determine the next child PIFO to schedule. Then, recursively inspect
the child PIFO to determine the next grand child PIFO to schedule,
until we reach a leaf PIFO that determines the next packet to
schedule.  Figure~\ref{fig:pifo_encoding} shows this encoding.

The instantaneous scheduling order of this PIFO tree can be modified as
packets are enqueued, by providing a scheduling transaction for each
node in the PIFO tree. This is our next programming abstraction: a
tree of such scheduling transactions.  Each node in this tree is a
tuple with two attributes (Figure~\ref{fig:hpfq_tree}). First, a
packet predicate specifies which packets execute that node's
scheduling transaction before the packet or a reference to another
PIFO is enqueued into that node's PIFO.  Second, a scheduling
transaction specifies how the rank is computed for elements
(packet or PIFO reference) that are enqueued into the node's PIFO.

When a packet is enqueued into a PIFO tree, it executes one transaction at each
node whose packet predicate matches the arriving packet. These nodes form a
path from a leaf to the root of the tree and the transaction at each node on
this path updates the scheduling order at that node. Notice that for each
packet, one element is enqueued into the PIFO at each node on the path from the
leaf to the root. At the leaf node, that element is the packet itself; at the
other nodes, it is a reference to another PIFO in the tree that eventually
points to the packet.  Packets are dequeued in the order encoded by the PIFO
tree~(Figure~\ref{fig:pifo_encoding}).  Figure~\ref{fig:hpfq} shows how a
network operator would program HPFQ using this tree abstraction.

\subsection{Shaping transactions}
\label{ss:hshaping}

\begin{figure*}
\begin{subfigure}[b]{0.2\textwidth}
\begin{center}
\includegraphics[width=\textwidth]{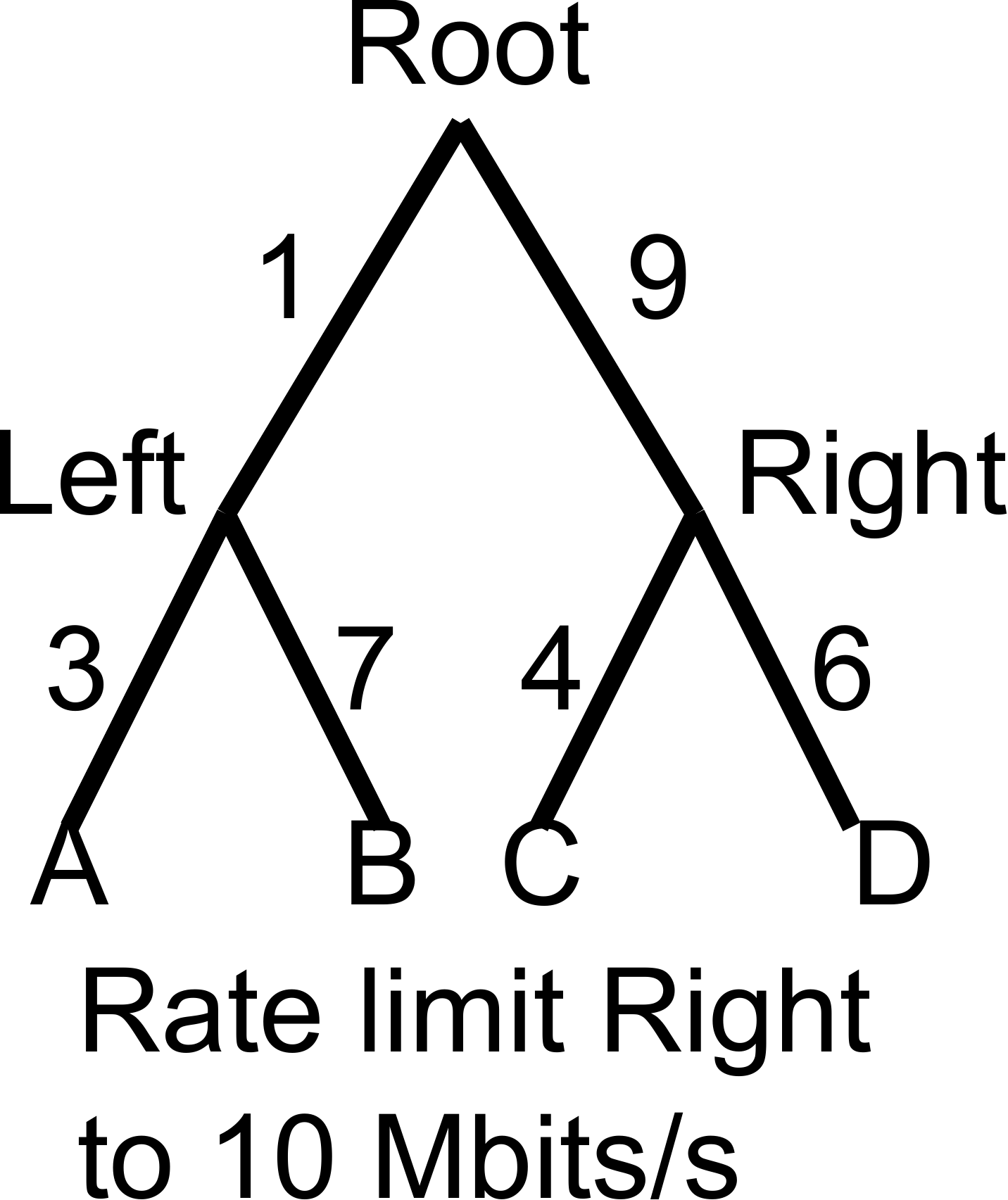}
\end{center}
\caption{Algorithm}
\label{fig:hshaping_algo}
\end{subfigure}
\vrule
\begin{subfigure}[b]{0.3\textwidth}
\begin{center}
\includegraphics[width=\textwidth]{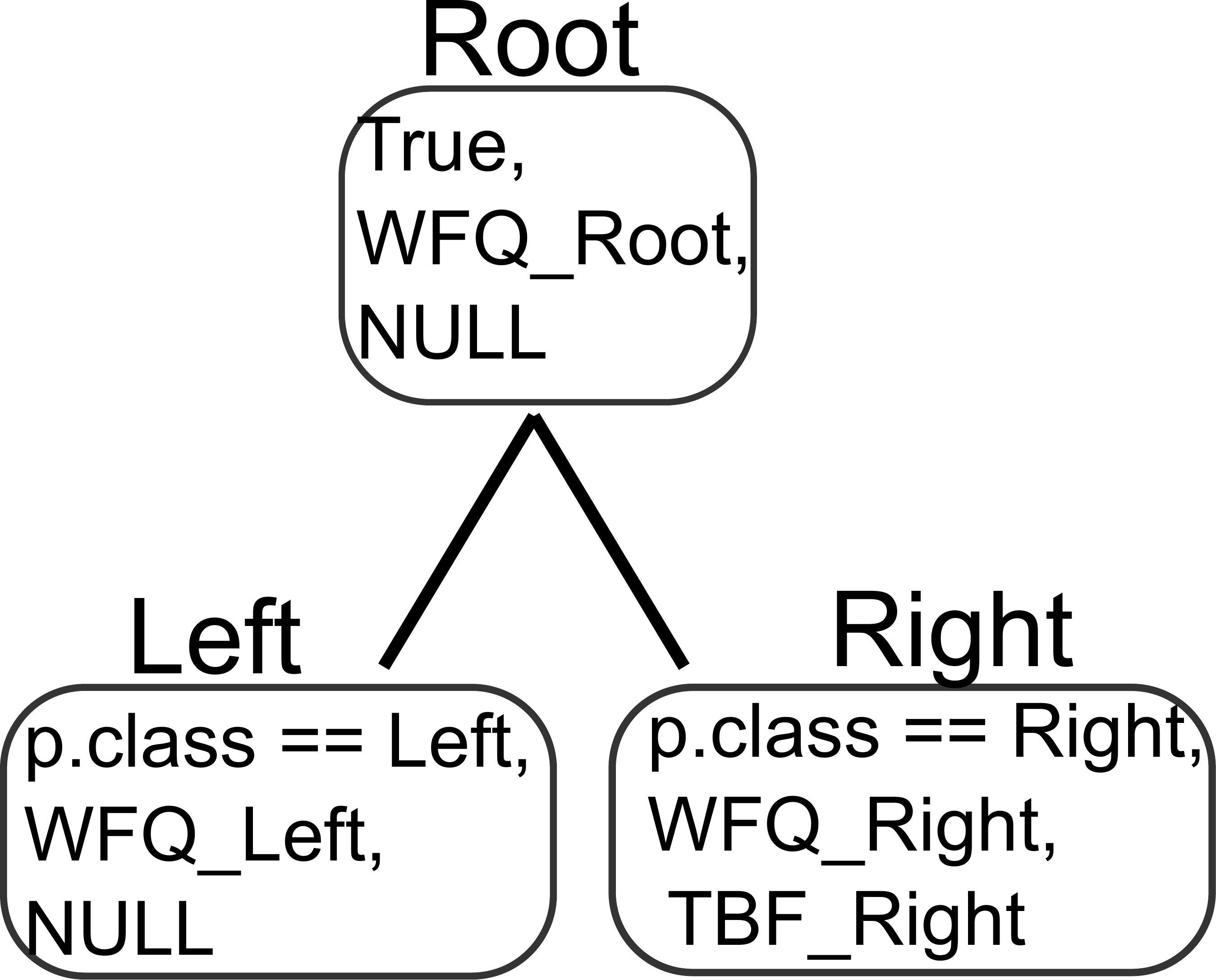}
\end{center}
\caption{Tree of PIFOs}
\label{fig:hshaping_tree}
\end{subfigure}
\vrule
\begin{subfigure}[b]{0.5\textwidth}
\begin{lstlisting}[style=customc]
tokens = min(tokens + r * (now - last_time), B)
if p.length <= tokens
  p.send_time = now
else
  p.send_time = now + (p.length - tokens) / r
tokens = tokens - p.length
last_time = now
p.rank = p.send_time
\end{lstlisting}
\caption{Shaping transaction for TBF\_Right.}
\label{fig:hshaping_shaping_trans}
\end{subfigure}
\caption{Programming Hierarchies with Shaping using PIFOs. The
  scheduling transactions for WFQ\_Right, WFQ\_Left, and WFQ\_Root are
  identical to Figure~\ref{fig:sched_trans}.}
\label{fig:hshaping}
\end{figure*}

So far, we have only considered work-conserving scheduling algorithms. Our
final abstraction, {\em shaping transactions}, allows us to program
non-work-conserving scheduling algorithms.

Non-work-conserving algorithms differ from work-conserving algorithms
in that they decide the {\em time} at which packets are scheduled as
opposed to just the scheduling {\em order}. As an example, consider
the scheduling algorithm shown in Figure~\ref{fig:hshaping_algo},
which extends the algorithm in Figure~\ref{fig:hpfq_algo} with the
additional requirement that the traffic in the Right class be limited
to 10 Mbit/s, regardless of the offered load. We refer to this example
throughout the paper as {\em Hierarchies with Shaping}.

To motivate our abstraction for this and other non-work-conserving algorithms,
recall that a PIFO tree encodes the instantaneous scheduling order, by walking
down the tree from the root PIFO to a leaf PIFO to schedule the next packet.
With this encoding, an element (packet or PIFO reference) can be scheduled only
if it resides in a PIFO and there is a chain of PIFO references from the root
PIFO to that element. To program non-work-conserving scheduling algorithms, we
provide the ability to defer when a packet or PIFO reference is enqueued into a
PIFO and hence is visible for scheduling.

To defer enqueues into PIFOs, we augment nodes of the tree with a {\em shaping
transaction} that is executed on all packets matched by the node's packet
predicate. The shaping transaction determines when a packet (or a reference to a
node's PIFO) is available for scheduling in the node's parent's PIFO. It
is only after the time set by the shaping transaction that the shaped packet or
PIFO reference is actually released to the parent node, where it is scheduled by
executing the parent's scheduling transaction and enqueuing it in its PIFO.

Figure~\ref{fig:hshaping_shaping_trans} shows an example of a shaping
transaction that implements a Token Bucket Filter (TBF) with a rate-limit of
$r$ and a burst allowance $B$.  Figure~\ref{fig:hshaping} shows how an operator
would use this shaping transaction to program Hierarchies with Shaping.  Here
the TBF shaping transaction (TBF\_Right) determines when the PIFO references
for class Right are released to its parent node (Root). Until PIFO references
for class Right are released to its parent, they are held in a separate
shaping PIFO, distinct from the node's scheduling PIFO. The shaping PIFO uses
the wall-clock departure time of an entry as its rank and pushes an entry to
the parent's scheduling PIFO when the entry's wall clock time arrives.

\begin{figure}[!h]
  \includegraphics[width=\columnwidth]{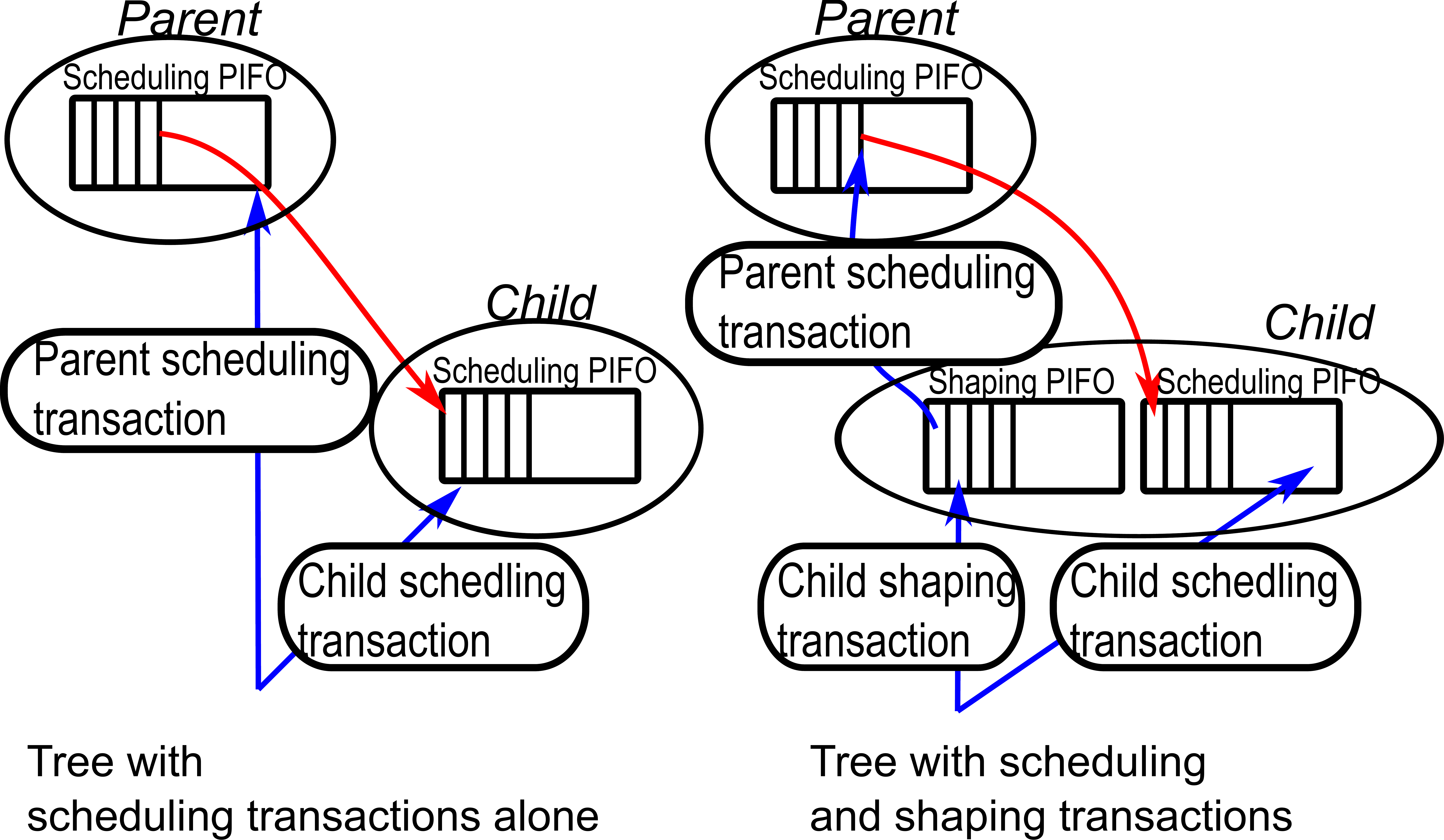}
  \caption{Relationship between scheduling and shaping transactions. The child
  shaping transaction suspends the execution of the parent scheduling transaction
  until the wall-clock time computed by the shaping transaction.}
  \label{fig:shaping_trans}
\end{figure}

\paragraph{The semantics of shaping transactions}
We explain the semantics of shaping transactions within a tree of scheduling
and shaping transactions using the two nodes $Child$ and $Parent$ shown in
Figure~\ref{fig:shaping_trans}. When a packet is enqueued, it executes a
scheduling transaction at the leaf node whose predicate matches this packet. It
then continues upward towards the root, as before, executing scheduling
transactions along the path, until it reaches the first node $Child$ that also
has a shaping transaction attached to it.

At this point, we execute two transactions at $Child$: the original scheduling
transaction to push an entry into $Child$'s scheduling PIFO and a shaping
transaction to push an element $R$, which is a reference to $Child$'s scheduling
PIFO, into $Child$'s shaping PIFO. After $R$ is pushed into $Child$'s shaping PIFO,
transactions are now suspended: no further transactions are executed when the
packet is enqueued.

Say $R$ had a wall-clock time $T$ as its rank when it was pushed into $Child$'s
shaping PIFO. At time $T$, $R$ will be dequeued from $Child$'s shaping PIFO and
enqueued into $Parent$'s scheduling PIFO, making $Child$'s scheduling PIFO now
visible to $Parent$. The rest of the path to the root is now resumed starting
at $Parent$. Note that this suspend-resume process can occur multiple times if
there are multiple nodes with shaping transactions along a packet's path from
its leaf to the root.  The dequeue logic remains identical to before, i.e.,
starting from the root dequeue recursively until we schedule a packet.

\section{The expressiveness of PIFOs}
\label{s:expressive}

In addition to the three detailed examples from \S\ref{s:pifo}, we now provide
three more examples (\S\ref{ss:lstf}, \S\ref{ss:stopngo}, \S\ref{ss:min_rate})
to demonstrate that our programming model built using the PIFO abstraction is
expressive. We also list several other examples that can be programmed using
PIFOs (\S\ref{ss:other}).

\subsection{Least Slack-Time First}
\label{ss:lstf}

Least Slack-Time First (LSTF)~\cite{lstf,ups} schedules packets at each switch
in increasing order of packet slacks, i.e., the time remaining until each
packet's deadline.  Packet slacks are initialized at an end host and
decremented by the wait time at each switch's queue. We can program LSTF at a
PIFO-enabled switch using the scheduling transaction in Figure~\ref{fig:lstf}.
To compute wait times at upstream switches, we tag packets with their
timestamps before and after they enter the queue, and compute the difference
between the two.\footnote{This can be achieved through proposals such as Tiny
Packet Programs~\cite{tpp} and In-Band Network Telemetry~\cite{int}.}

\begin{figure}
\begin{lstlisting}[style=customc]
  p.slack = p.slack - p.prev_wait_time;
  p.rank  = p.slack;
\end{lstlisting}
\caption{Scheduling transaction for LSTF}
\label{fig:lstf}
\end{figure}

\subsection{Stop-and-Go Queueing}
\label{ss:stopngo}

Stop-and-Go Queueing~\cite{stopngo} is a non-work-conserving algorithm
that provides bounded delays to packets using a framing strategy. Time
is divided into non-overlapping frames of equal length $T$, where
every packet arriving within a frame is transmitted at the end of the
frame, smoothing out any burstiness in traffic patterns induced by
previous hops. To program Stop-and-Go Queueing, we use the shaping
transaction in Figure~\ref{fig:stopngo}.  {\tt frame\_begin\_time} and
{\tt frame\_end\_time} are two state variables that track the
beginning and end of the current frame (in wall-clock time).  When a
packet is enqueued, its departure time is set to the end of the
current frame. Multiple packets with the same departure time are sent
out in first-in first-out order, as guaranteed by a PIFO's semantics
(\S\ref{s:pifo}).

\begin{figure}
  \begin{lstlisting}[style=customc]
  if (now >= frame_end_time):
    frame_begin_time = frame_end_time
    frame_end_time   = frame_begin_time + T
  p.rank = frame_end_time
  \end{lstlisting}
\caption{Shaping transaction for Stop-and-Go Queueing}
\label{fig:stopngo}
\end{figure}

\subsection{Minimum rate guarantees}
\label{ss:min_rate}

A common scheduling policy on many switches today is providing a minimum rate
guarantee to a flow, provided the sum of all such guarantees doesn't exceed the
link capacity. A minimum rate guarantee can be programmed using PIFOs by using
a two-level PIFO tree, where the root of the tree implements strict priority
scheduling across flows: those flows below their minimum rate are scheduled
preferentially to flows above their minimum rate. Then, at the next level of
the tree, flows implement the first-in first-out discipline across their packets.

To program minimum rate guarantees, when a packet is enqueued, it executes a
FIFO scheduling transaction at its leaf node, setting its rank to the
wall-clock time on arrival. At the root level, a PIFO reference (the packet's
flow identifier) is pushed into the root PIFO using a rank that reflects
whether the flow is above or below its rate limit after the arrival of the
current packet. To determine this, we run the scheduling transaction in
Figure~\ref{fig:min_rate} that uses a token bucket (tb) that can be filled up
until BURST\_SIZE to decide if the arriving packet puts the flow below or above
a particular flow rate.

\begin{figure}
  \begin{lstlisting}[style=customc]
  // Replenish tokens
  tb = tb + min_rate * (now - last_time);
  if (tb > BURST_SIZE) tb = BURST_SIZE;

  // Check if we have enough tokens
  if (tb > p.size):
    p.over_min = 0; // under min. rate
    tb = tb - p.size;
  else:
    p.over_min = 1; // over min. rate

  last_time = now;
  p.rank = p.over_min;
  \end{lstlisting}
\caption{Scheduling transaction for min. rate guarantees.}
\label{fig:min_rate}
\end{figure}

Note that ``collapsing'' this tree into a single node that implements the
scheduling transaction in Figure~\ref{fig:min_rate} does not work because it
causes packet reordering within a flow: an arriving packet can cause a flow to
move from a lower to a higher priority and, in the process, leave before
low priority packets from the flow that arrived earlier. The 2-level tree solves this problem by
attaching priorities to transmission opportunities for a specific flow and not
specific packets. Now if an arriving packet causes a flow to move from low to
high priority, the next packet that is scheduled from this flow is the earliest
packet from that flow---not the arriving packet.

\subsection{Other examples}
\label{ss:other}

We now briefly describe several more examples of scheduling algorithms that can
be programmed using PIFOs.

\begin{CompactEnumerate}
  \item \textbf{Fine-grained priority scheduling:} Many algorithms implement
    fine-grained priority scheduling and schedule the packet with the lowest
    value of a field initialized by the end host. These algorithms can be
    programmed by using a scheduling transaction to set the packet's rank to
    the appropriate field. Examples of such algorithms and the fields they use
    are strict priority scheduling (IP TOS field), Shortest Job First (flow
    size), Shortest Remaining Processing Time (remaining flow size), Least
    Attained Service (service received in bits for a flow), and Earliest
    Deadline First (time until a deadline).
  \item \textbf{Service-Curve Earliest Deadline First (SC-EDF)~\cite{sced}} is
    a scheduling algorithm that schedules packets in increasing order of
    a deadline computed from a flow's service curve (a specification of the
    service a flow should receive over any given time interval). We
    can program SC-EDF using a scheduling transaction that sets a packet's rank
    to the deadline computed by the SC-EDF algorithm.
   \item \textbf{First-In First-Out} can be programmed by using a scheduling
     transaction that sets a packet's rank to the wall-clock time on arrival.
   \item \textbf{Rate-Controlled Service Disciplines (RCSD)~\cite{rcsd}} are a
     class of non-work-conserving scheduling algorithms that can be implemented
     using a combination of a rate regulator to shape traffic and a packet
     scheduler to schedule traffic. An algorithm from the RCSD framework can be
     programmed using PIFOs by expressing the rate regulator using a shaping
     transaction and the packet scheduler using a scheduling transaction.
     Examples in this class include Jitter-EDD~\cite{jitteredd} and Hierarchical Round Robin~\cite{hrr}.
   \item \textbf{Class-Based Queueing (CBQ)~\cite{cbq, cbq_impl}} is a
     hierarchical scheduling algorithm that first schedules among classes based
     on a priority field assigned to each class, and then uses fair queueing to
     schedule packets within a class. CBQ can be programmed by a using a
     two-level PIFO tree to realize inter-class and intra-class scheduling.
\end{CompactEnumerate}

\subsection{Limitations}
\label{ss:limitations}

We conclude by describing some algorithms that cannot be programmed using the
PIFO abstraction.

\paragraph{Changing the scheduling order of all packets in a flow}

A PIFO allows an arriving element (packet or PIFO reference) to determining its
own scheduling order relative to all other elements currently in the PIFO. A
PIFO does not allow the arriving element to change the scheduling order of all
elements belonging to that element's flow that are already present in the PIFO.

An example of an algorithm that needs this capability is
pFabric~\cite{pFabric}, which schedules the earliest packet from the flow with
the shortest remaining processing time, to prevent packet reordering. To
illustrate why this is beyond a PIFO's capabilities, consider the sequence of
arrivals below, where pi(j) represents a packet from flow i with remaining
processing time j.
\begin{CompactEnumerate}
\item Enqueue p0(7).
\item Enqueue p1(9), p1(8).
\item The departure order now is: p0(7), p1(9), p1(8).
\item Enqueue p1(6).
\item The departure order now is: p1(9), p1(8), p1(6), p0(7).
\end{CompactEnumerate}

The order of all packets from flow 1 needs to change in response to the arrival
of a single packet p1(6) from flow 1, which is beyond a PIFO's capabilities.

\paragraph{Traffic shaping across multiple nodes in a tree}

Our programming model for scheduling attaches a single shaping and scheduling
transaction to a node. This lets us enforce rate limits on a single node, but
not across multiple nodes in the tree. As an illustration, PIFOs cannot express
the following policy: WFQ on a set of flows A, B, and C, with the additional
constraint that the aggregate throughput of A and B together doesn't exceed 10
Mbit/s. One work around is to implement this as HPFQ across two classes C1 and
C2, with C1 containing A and B, and C2 containing C alone. Then, we enforce the
rate limit of 10 Mbit/s on C1. However, this isn't equivalent to our desired
policy. More generally, our programming model for programmable scheduling
establishes a 1-to-1 relationship between the scheduling and shaping
transactions, which is constraining for some algorithms.

\paragraph{Output rate limiting}

The PIFO abstraction enforces rate limits using a shaping transaction, which
determines a packet or PIFO reference's scheduling time before it is enqueued
into a PIFO.  The shaping transaction permits rate limiting on the {\em input}
side, i.e., before elements are enqueued. An alternate form of rate limiting is
on the {\em output}, i.e., by limiting the rate at which elements are scheduled.

As an example, consider a scheduling algorithm with two priority queues, low
and high, where low is to be rate limited to 10 Mbit/s. To program this using
input side rate limiting, we would use a shaping transaction to impose a 10
Mbit/s rate limit on low and a scheduling transaction to implement strict
priority scheduling between low and high. Now, assume packets from high starve
low for an extended period of time. During this time, packets from low get rate
limited through the shaping transaction and accumulate in the PIFO shared with
high. Now, if there are suddenly no more high packets, all packets from low
would get scheduled out at line rate, and no longer be rate limited to 10
Mbit/s over a transient period of time (until all instances of low are drained
out of the PIFO shared with high). Input rate limiting still provides long-term
guarantees on the rate, while output rate limiting provides these guarantees on
shorter time scales as well.

\section{Design}
\label{s:design}

We now present a hardware design for a programmable scheduler based on
PIFOs. For concreteness, we target line-rate switches with an
aggregate capacity of up to a billion packets/s, i.e., a 1 GHz clock
frequency for the switch pipeline. Examples of such switch
architectures include the RMT architecture~\cite{rmt} and Intel's
FlexPipe~\cite{flexpipe}, both of which provide 64 10 Gbit/s ports,
which at the minimum packet size of 64 bytes translates to a billion
packets/s.

For our hardware design, we first describe how scheduling and shaping
transactions can be implemented (\S\ref{ss:transactions}). Then, we
show how a tree of PIFOs can be realized using a full mesh of PIFO
blocks by appropriately interconnecting these blocks
(\S\ref{ss:mesh}). We also describe how a compiler
(\S\ref{ss:compiler}) can automatically configure this mesh given the
description of the scheduling algorithm as a tree of scheduling and
shaping transactions.

\subsection{Scheduling and shaping transactions}
\label{ss:transactions}

Scheduling and shaping transactions compute an element's rank and execute
atomically. By this, we mean that the state of the system (both the PIFO and
any auxiliary state used in the transaction) after $N$ enqueues is equivalent
to serially executing $N$ transactions one after the other, with no overlap
between them. For our purpose, we need to execute these transactions atomically
at line rate, i.e., the rate at which the switch receives packets (e.g., a
billion packets/s).

To implement scheduling and shaping transactions, we use
Domino~\cite{domino_arxiv}, a recent system to program data-plane
algorithms at line rate using packet transactions. Domino extends work
on programmable line-rate switches~\cite{rmt, flexpipe, xpliant} by
providing hardware primitives, called atoms, and software
abstractions, called packet transactions, to program data-plane
algorithms at line rate. Atoms are small processing units that
represent a programmable switch's instruction set. The Domino compiler
compiles a packet transaction into a pipeline of atoms that executes
the transaction atomically, rejecting the packet transaction if it
can't run at line rate. Scheduling and shaping transactions are, in
fact, packet transactions written in the Domino language.

The Domino paper proposes atoms that are rich enough to support many
data-plane algorithms and small enough that they can be implemented at
1 GHz with modest chip area overhead. The largest of these atoms,
called Pairs, occupies an area of 6000 \si{\micro\metre\squared} in a
32 nm standard-cell library; a switch with a chip area of 200
\si{\milli\metre\squared}~\cite{glen_parsing} can support 300 of these
with less than 1\% area overhead. The Domino paper further shows how
the transaction in Figure~\ref{fig:sched_trans} can be run at 1 GHz on
a switch pipeline with the Pairs atom.

In a similar manner, we could use the Domino compiler to compile scheduling and
shaping transactions to a pipeline of atoms for other scheduling and shaping
transactions. For example, the transactions for Token Bucker Filtering (Figure
~\ref{fig:hshaping_shaping_trans}), minimum rate guarantees
(Figure~\ref{fig:min_rate}), Stop-and-Go queueing(Figure~\ref{fig:stopngo}),
and LSTF (Figure~\ref{fig:lstf}), can all be expressed as Domino programs.

\subsection{The PIFO mesh}
\label{ss:mesh}
\begin{figure}
  \centering
  \includegraphics[width=0.65\columnwidth]{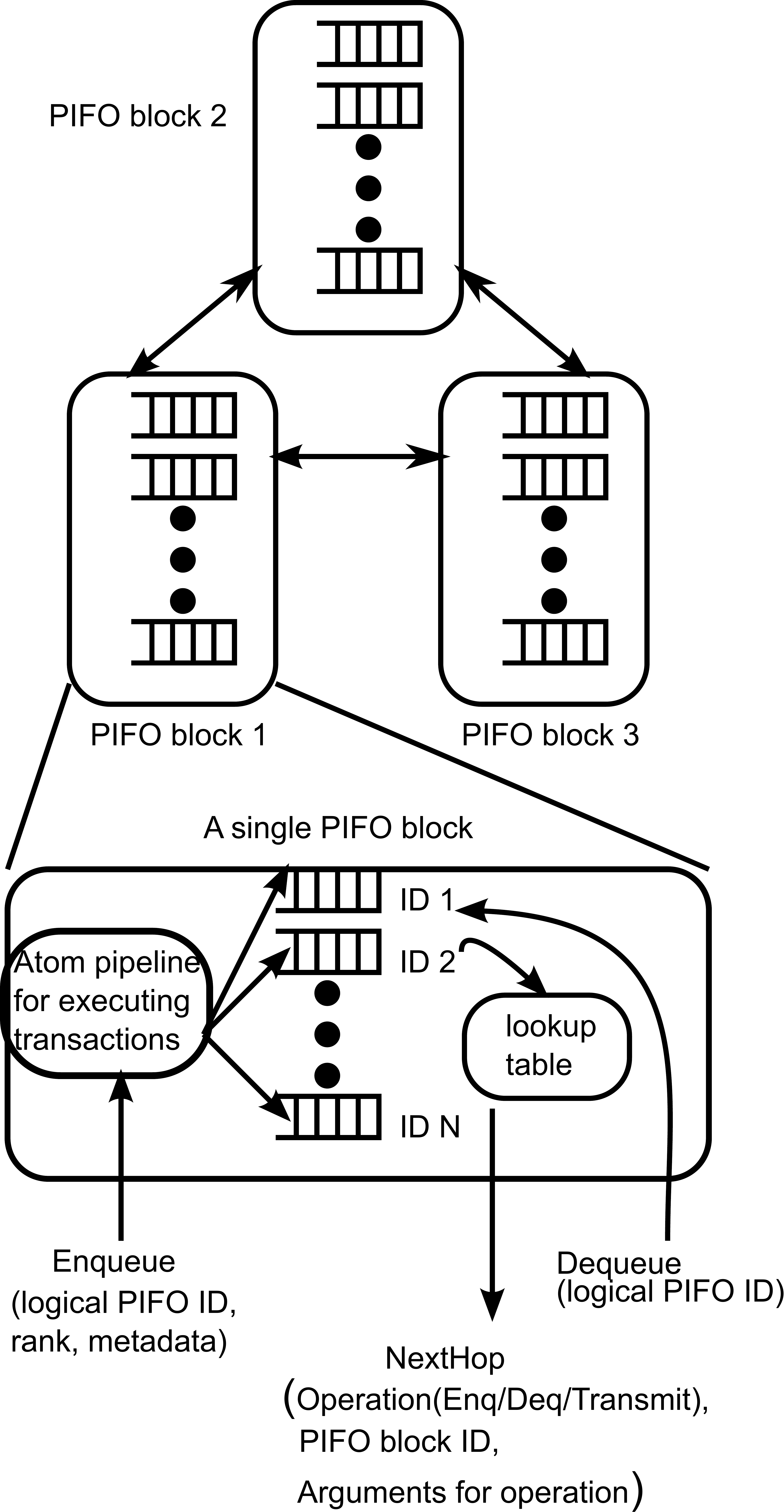}
  \caption{Block diagram of PIFO mesh}
  \label{fig:mesh}
\end{figure}

The next component of the hardware is the actual PIFO itself. We lay out PIFOs
physically as a full mesh of {\em PIFO blocks} (Figure~\ref{fig:mesh}), where
each block implements multiple logical PIFOs. We expect a small number of PIFO
blocks in a typical switch (e.g., less than five). The PIFO blocks correspond
to different levels of a hierarchical scheduling tree. Most practical
scheduling algorithms do not require more than a few levels of hierarchy,
implying the required number of PIFO blocks is small as well. As a result, a
full mesh between these blocks is feasible (see ~\S\ref{ss:interconnect} for
more details).

Each PIFO block runs at a clock frequency of 1 GHz and contains an atom
pipeline to execute scheduling and shaping transactions. In every clock cycle,
each PIFO block supports one enqueue and dequeue operation into any one
of the logical PIFOs residing within that block. We address a logical PIFO
within a block with a logical PIFO ID.

The interface to a PIFO block is:
\begin{CompactEnumerate}
\item Enqueue an element (packet or reference to another PIFO) given a
  logical PIFO ID, the element's rank, and some metadata that will be
  carried with the element (such as the packet length required for
  STFQ's rank computation). The enqueue returns nothing.
  \item Dequeue from a specific logical PIFO ID within the block. The dequeue
    returns either a packet or a reference to another PIFO.
\end{CompactEnumerate}

After a dequeue, besides transmitting a packet, a PIFO block may need to
communicate with another PIFO block for one of two reasons:
 \begin{CompactEnumerate}
 \item To dequeue a logical PIFO in another block (e.g., when
   dequeuing a sequence of PIFOs from the root to a leaf of a
   scheduling tree to transmit packet).
 
 \item To enqueue into a logical PIFO in another block (e.g., when
   enqueuing a packet that has just been dequeued from a shaping
   PIFO).
 \end{CompactEnumerate}

We configure these post-dequeue operations using a small lookup table, which
looks up the ``next hop'' following a dequeue. This lookup table specifies an
operation (enqueue, dequeue, transmit), the PIFO block for the next operation,
and any arguments the operation needs.

\subsection{Compiling to a PIFO mesh}
\label{ss:compiler}

\begin{figure*}[!t]
  \begin{subfigure}[b]{0.5\textwidth}
  \begin{center}
  \includegraphics[width=0.8\textwidth]{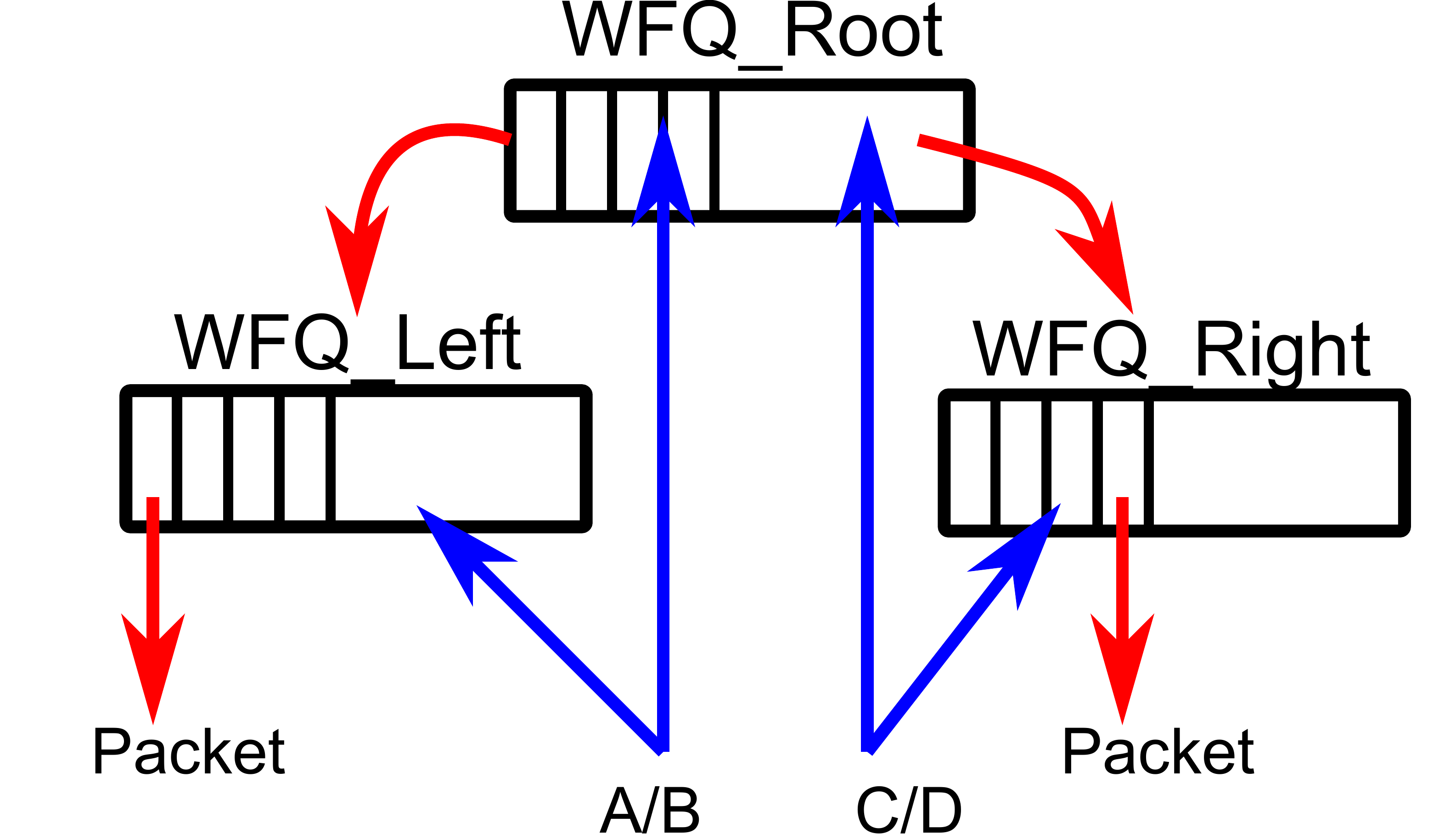}
  \caption{Enqueue and dequeue paths for HPFQ. Downward arrows indicate dequeue
  paths.  Upward arrows indicate enqueue paths.}
  \label{fig:hpfq_path}
  \end{center}
  \end{subfigure}
  \begin{subfigure}[b]{0.5\textwidth}
  \begin{center}
  \includegraphics[width=0.8\textwidth]{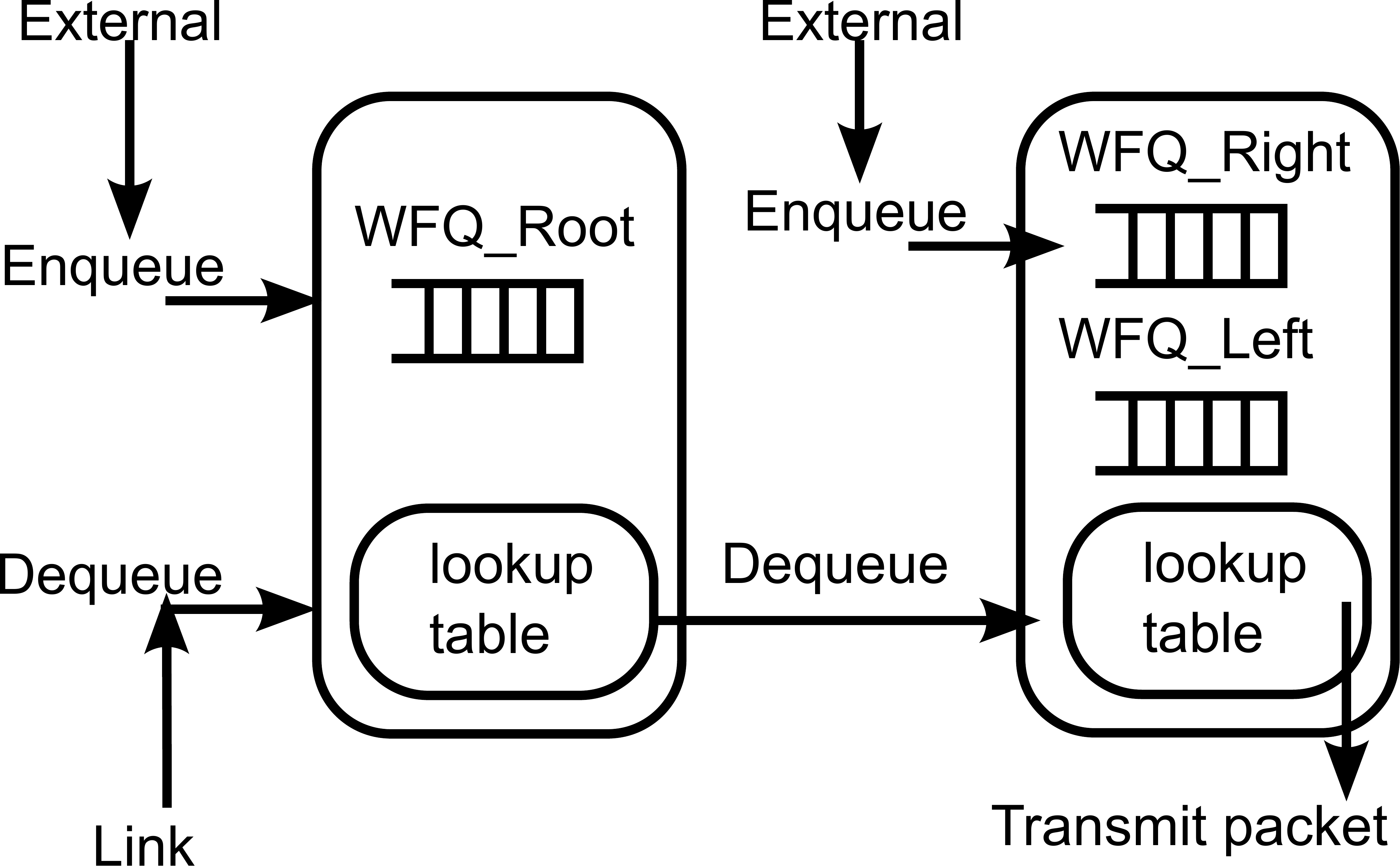}
  \caption{Mesh configuration for HPFQ.}
  \label{fig:hpfq_mesh}
  \end{center}
  \end{subfigure}
  \caption{Compiling HPFQ to a PIFO mesh}
\end{figure*}

\begin{figure*}[!t]
  \begin{subfigure}[b]{0.5\textwidth}
  \includegraphics[width=\textwidth]{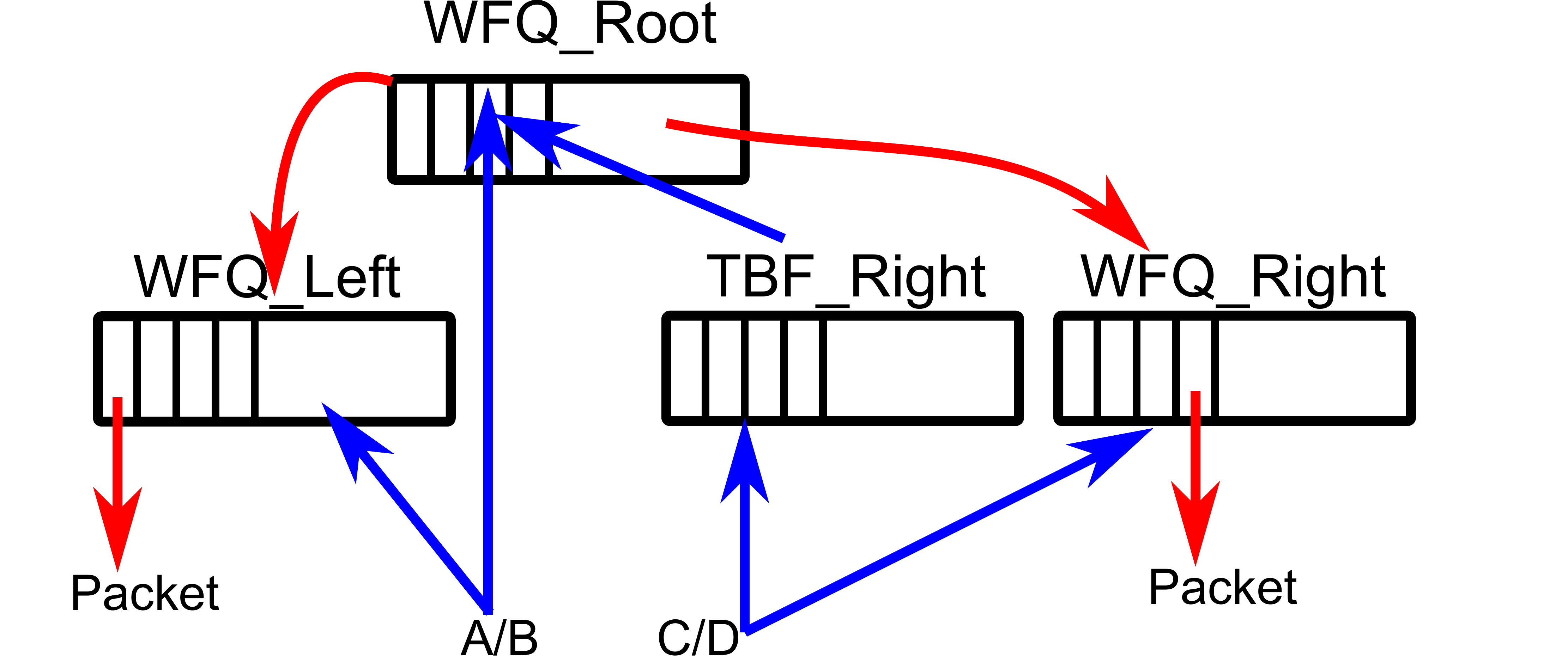}
  \caption{Enqueue and dequeue paths for hierarchies with shaping. Downward
  arrows indicate dequeue paths. Upward arrows indicate enqueue paths.}
  \label{fig:hshaping_path}
  \end{subfigure}
  \begin{subfigure}[b]{0.5\textwidth}
  \includegraphics[width=\textwidth]{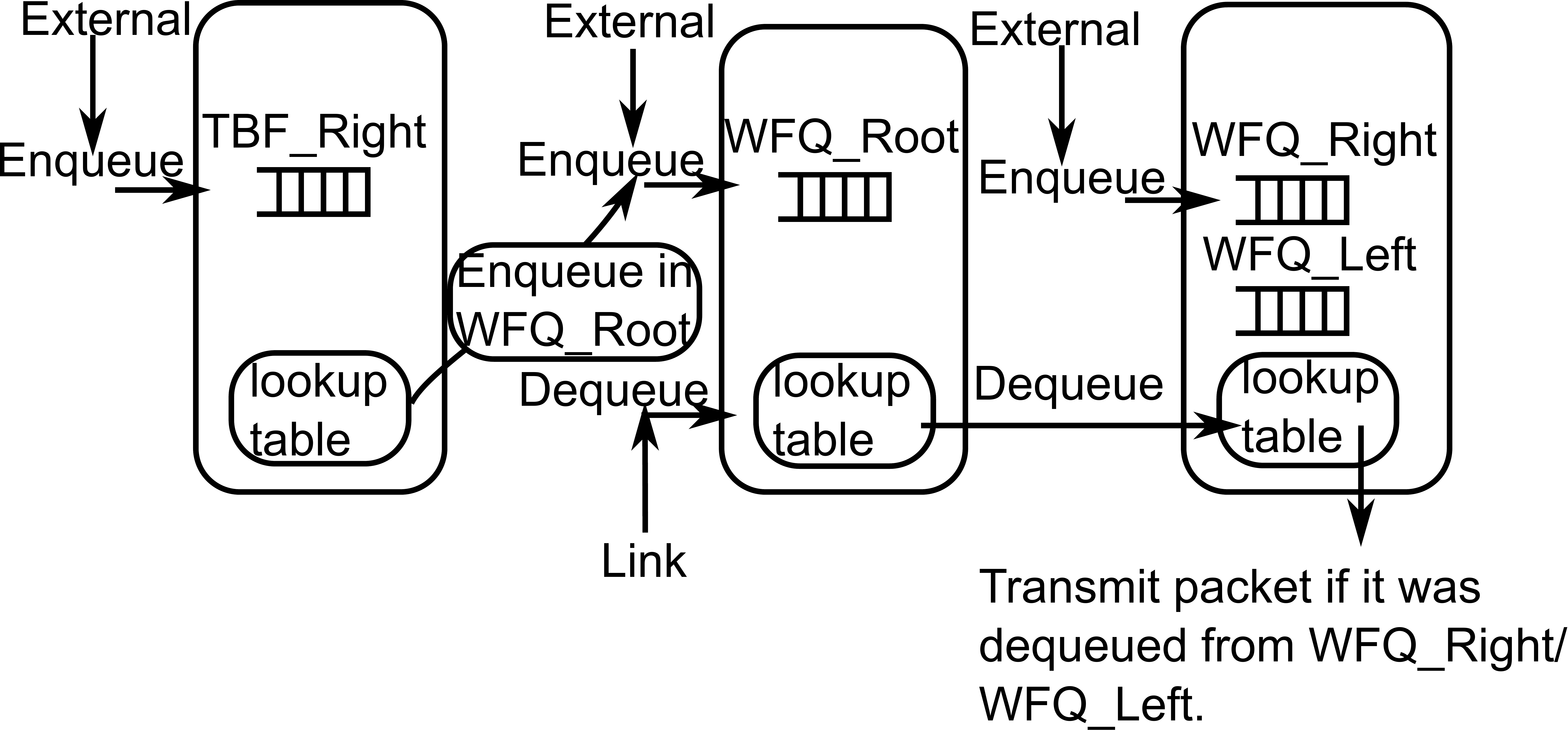}
  \caption{Mesh configuration for Hierarchies with Shaping.}
  \label{fig:hshaping_mesh}
  \end{subfigure}
  \caption{Compiling Hierarchies with Shaping to a PIFO mesh}
\end{figure*}

A PIFO mesh is configured by specifying the logical PIFOs within each PIFO
block and by populating each PIFO block's next-hop lookup table. A compiler
could configure the PIFO mesh from the scheduling algorithm specified as a tree
of scheduling and shaping transactions. We illustrate the compilation using
examples from Figures~\ref{fig:hpfq} and \ref{fig:hshaping}.  The compiler
first converts the tree with scheduling and shaping transactions to an
equivalent tree representation that specifies the enqueue and dequeues
operations on each PIFO.  Figures~\ref{fig:hpfq_path} and
\ref{fig:hshaping_path} show this representation for Figures~\ref{fig:hpfq} and
\ref{fig:hshaping} respectively.

It then overlays this tree over a PIFO mesh by assigning every level of the
tree to a PIFO block and configuring the lookup tables to connect PIFO blocks
as required by the tree.  Figure~\ref{fig:hpfq_mesh} shows the PIFO mesh
corresponding to Figure~\ref{fig:hpfq}. If a particular level of the tree has
more than one enqueue or dequeue from another level, we allocate new PIFO
blocks as required to respect the constraint that any PIFO block provides one
enqueue and dequeue operation per clock cycle. This is shown in
Figure~\ref{fig:hshaping_mesh}, which has an additional PIFO block containing
TBF\_Right\footnote{More precisely, the shaping PIFO that the TBF\_Right transaction
enqueues into. We use a transaction's name to refer both to the transaction and the
PIFO it enqueues into.} alone.

\paragraph{Challenges with shaping transactions}

Each PIFO block supports one enqueue and dequeue operation per clock
cycle.  This suffices to implement any algorithm which only uses
scheduling transactions (i.e. work-conserving algorithms) at
line-rate. The reason is that for such algorithms each packet needs at
most one enqueue and one dequeue at each level of its scheduling tree,
and we map the PIFOs at each level to a different PIFO block.

However, shaping transactions pose challenges. Consider the
non-work-conserving algorithms in Figure~\ref{fig:hshaping}. When the
shaping transaction enqueues elements into TBF\_Right, these elements
will be released into WFQ\_Root at a future time $T$. The external
enqueue into WFQ\_Root may also happen exactly at $T$, because a
packet arrives at that time. This creates a conflict because there are
two enqueue operations in the same cycle.  Conflicts may also manifest
on the dequeue side, e.g., if TBF\_Right shared its PIFO block with
another logical PIFO, dequeue operations to the two logical PIFOs
could occur at the same time because TBF\_Right can be dequeued at any
arbitrary wall-clock time.

In a conflict, only one of the two operations can proceed. We resolve
this conflict in favor of PIFOs where element ranks are computed by
scheduling transactions. This reflects the intuition that PIFOs
controlled by shaping transactions are used for rate limiting to a
rate lower than the line rate at which packets are normally
scheduled. As a result, they can afford to be delayed by a few clocks
until there are no more conflicts. By contrast, delaying scheduling
decisions of a PIFO controlled by a scheduling transaction would mean
that the switch would idle without transmitting a packet and not
satisfy its line-rate guarantee.

Effectively, this means that PIFOs controlled by shaping transactions
only get best-effort service. There are workarounds to this
undesirable situation. One is to overclock the pipeline at a higher
clock rate than required for packet processing, such as 1.25 GHz
instead of 1 GHz, providing a few spare clock cycles for best-effort
processing. Another is to provide multiple ports to a PIFO block to
support multiple operations every clock. These techniques are commonly
used in switches today for low priority background tasks such as
reclaiming buffer space.  They can be applied to the PIFO mesh as
well.

\section{Hardware Implementation}
\label{s:hardware}

We now describe the detailed implementation of a PIFO mesh. First, we discuss
the implementation of a single PIFO block within the mesh
(\S\ref{ss:single_block}).  Then, we synthesize this implementation to a 16 nm
standard-cell library and evaluate its area overheads (\S\ref{ss:feasibility}).
Finally, we evaluate the feasibility of interconnecting these PIFO blocks using
a full mesh (\S\ref{ss:interconnect}).

\subsection{Performance requirements}
\label{ss:performance}

Our goal in implementing a PIFO block is to build a scheduler that is
performance-competitive with current shared-memory switches, such as the
Broadcom Trident~\cite{trident}, which are commonly used in datacenters today.
As concrete performance targets, we target a 1 GHz clock frequency, which
supports 64 10 Gbit/s ports. Based on the Broadcom Trident, we target a packet
buffer size of 12 MBytes, and a cell size\footnote{Packets in a shared-memory
switch are allocated in small units called cells.} of 200 bytes. In the worst
case, every packet is a single cell, implying the need to support up to 60K
packets/elements per PIFO block spread over multiple logical PIFOs.
Similarly, based on the Broadcom Trident, we set a target of 1000 flows over
which scheduling decisions are made at any port.

\subsection{A single PIFO block}
\label{ss:single_block}

Functionally, a single PIFO block needs to support two operations: an enqueue
operation that inserts an element into a logical PIFO and a dequeue operation
to dequeue from the head of a logical PIFO.  We first describe an
implementation for a single logical PIFO and then extend it to multiple logical
PIFOs in the same physical PIFO block.

A naive implementation is a flat sorted array of elements. An incoming element
is compared against all elements in parallel to determine a location for the
new element. It is then inserted into this location by shifting the array.
However, each comparison needs an independent comparator circuit and supporting
~60K of these is infeasible.

\begin{figure*}
  \centering
  \includegraphics[width=0.8\textwidth]{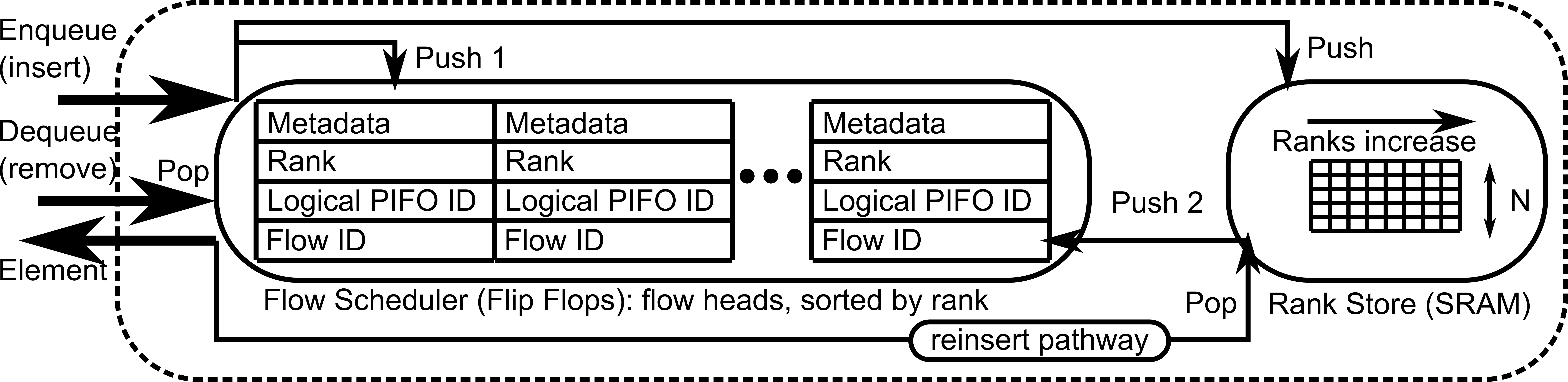}
  \caption{A single PIFO block with a flow scheduler and a rank
    store.}
  \label{fig:pifo_block}
\end{figure*}

However, nearly all practical scheduling algorithms naturally group packets
into flows or classes (e.g., based on traffic type, ports, or addresses)
and schedule packets of a flow in FIFO order. In these algorithms, packet ranks
strictly increase across consecutive packets in a flow.  This motivates a PIFO
design (Figure~\ref{fig:pifo_block}) with two parts: (1) a {\em flow scheduler}
that picks the element to dequeue based on the rank of the {\em head} element
of each flow, i.e., the element that arrived earliest, and (2) a {\em rank store}, a
bank of FIFOs that stores element ranks beyond the head elements.  This
decomposition reduces the number of elements that need sorting from the
number of packets (60 K) to the number of flows (1K). During an
enqueue, an element (both rank and metadata) is appended to the end of the
appropriate FIFO in the rank store.\footnote{If this is the first element in
  the flow, it bypasses the rank store and is directly pushed into the flow
scheduler data structure.} To permit enqueues into this PIFO block, we also
supply a flow ID argument to the enqueue operation.

Besides better scalability, an added benefit of this design is the ability to
reuse a significant amount of engineering effort that has gone into building
hardware IP for a bank of FIFOs. In a FIFO bank, each FIFO can grow and shrink
dynamically as required, subject to an overall limit on the number of entries
across the bank.  Such banks are commonly used today to buffer packet data in a
switch scheduler. As a result, we focus our implementation effort on building
the flow scheduler alone.

\paragraph{The Flow Scheduler}

The core operation within the flow scheduler is to sort an array of
flows based on the ranks of flow heads. The flow scheduler needs to
support one enqueue and one dequeue to the PIFO block every clock
cycle, which translates into two operations on the flow scheduler
every clock cycle:
\begin{CompactEnumerate}
  \item Inserting a flow when the flow goes from empty to non-empty (for the enqueue operation).
  \item Removing a flow that goes empty once it is scheduled,
        or reinserting a flow with the priority of the next element if it is still backlogged once it is scheduled (for the dequeue operation).
\end{CompactEnumerate}

The operations above require the ability to push up to two elements
into the flow scheduler every cycle (one each for the insert and
reinsert) and the ability to pop one element every cycle (for either
the remove or reinsert). These operations
require parallel operations on all elements in the flow scheduler. To
facilitate this, we implement the flow scheduler data structure in
flip flops (unlike the rank store, which is stored in SRAM).

Internally, the flow scheduler is organized as a sorted array, where a
push is implemented by:
\begin{CompactEnumerate}
\item Comparing against all elements in the array in parallel, using a comparator
  circuit, to produce a bit mask with the comparison results.
\item Finding the first 0-1 transition in the bit mask, using a priority encoder
  circuit, to determine the index to push into.
\item Pushing the element into the appropriate index, by shifting the array.
\end{CompactEnumerate}
A pop is implemented by shifting the head element out of the sorted array.

So far, we have focused on the flow scheduler for a single logical PIFO. To
handle multiple logical PIFOs, we keep elements sorted by rank, regardless of
which logical PIFO they belong to; hence, the push implementation doesn't
change. To pop from a logical PIFO, we compare against all elements to
determine elements with the same logical PIFO ID. Among these, we find the
first using a priority encoder. We then remove this element by shifting the
array.

\begin{figure}
  \includegraphics[width=0.5\textwidth]{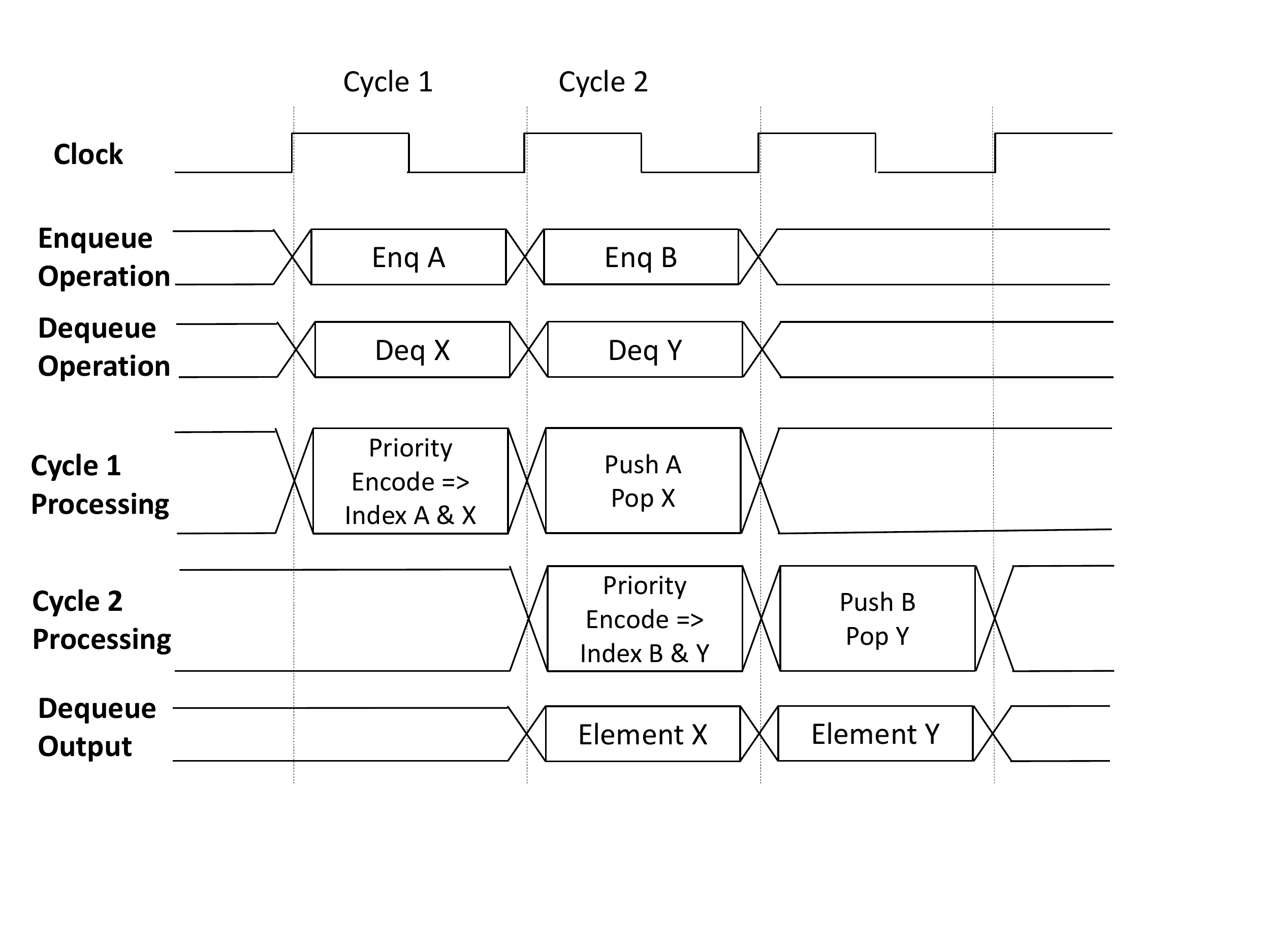}
  \caption{2 stage pipeline for flow scheduler}
  \label{fig:2stage}
\end{figure}

The internal push and pop operations require 2 clock cycles each and hence need
to be pipelined to support a throughput of 2 pushes and 1 pop every clock
cycle. The first stage of this 2-stage pipeline for the push operation
carries out the parallel comparison and priority encoder steps to determine an
index; the second stage actually pushes the element into the array using
the index. Similarly, for the pop operation, the first stage carries out the
equality check (for logical PIFO IDs) and priority encoder steps to compute an
index; the second stage pops the head element out of the array using the
index.  Figure~\ref{fig:2stage} shows the 2-stage pipeline.

The pipelined implementation meets timing at 1 GHz and supports up to
one enqueue/dequeue operation to any logical PIFO within a PIFO block
every clock cycle. Because pops take 2 cycles, and a reinsert
operation requires a pop followed by an access to the rank store for the
next element, followed by a push, our implementation supports a
dequeue from the same logical PIFO only once every 3 cycles (2 cycles
for the pop and 1 cycle to access the rank store in SRAM). This
restriction is inconsequential in practice. A dequeue every 3 cycles
from a logical PIFO is sufficient to service the highest link speed in
current switches, 100 Gbit/s, which requires a dequeue at most once every
5 clock cycles for a minimum packet size of 64 bytes. Dequeues to
distinct logical PIFO IDs are still permitted every cycle.

\subsection{Area overhead}
\label{ss:feasibility}

We synthesized the design described above to a gate-level netlist in a
16-nm standard cell library to determine its area overhead. We first
calculate area overheads for a baseline PIFO block that supports 1024
flows that can be shared across 256 logical PIFOs, and uses  a 16-bit
rank field and a 32-bit metadata field for each element in the PIFO.
In addition, we assume our rank store supports 64K elements.

Table~\ref{tab:area_overheads} calculates chip-area overheads when synthesized
to a 16-nm standard-cell library. Overall, a 5-block PIFO mesh consumes about
7.35 \si{\milli\metre\squared} of chip area (including the area of the atom
pipelines for rank computations). This is about 3.7\% of the chip area of a
typical switching chip today (using the minimum chip area estimate of 200
\si{\milli\metre\squared} provided by Gibb et al.~\cite{glen_parsing}). Of
course, a 5-block PIFO mesh is a significantly more capable packet scheduler
compared to current switch schedulers.
 
\begin{table}[!h]
  \begin{small}
  \begin{tabular}{|p{0.18\textwidth}|p{0.25\textwidth}|}
  \hline
  Component & Area in \si{\milli\metre\squared}\\
  \hline
  Switching chip & 200--400~\cite{glen_parsing} \\
  \hline
  Flow Scheduler & 0.224 (from synthesis) \\
  \hline
  SRAM (1 Mbit) & 0.145~\cite{sram_estimate} \\
  \hline
  Rank store & 64 K * (16 + 32) bits * 0.145 \si{\milli\metre\squared} / Mbit = 0.445 \\
  \hline
  Next pointers for linked lists in dynamically allocated rank store & 64 K * 16 bit pointers * 0.145 = 0.148 \\
  \hline
  Free list memory for dynamically allocated rank store & 64 K * 16 bit pointers * 0.145 = 0.148 \\
  \hline
  Head, tail, and count memory for each flow in the rank store & 0.1476 (from synthesis) \\
  \hline
  One PIFO block & 0.224 + 0.445 + 0.148 + 0.148 + 0.1476 = 1.11 \si{\milli\metre\squared} \\
  \hline
  5-block PIFO mesh & 5.55 \\
  \hline
  300 atoms spread out over the 5-block PIFO mesh for rank computations & 6000 \si{\micro\metre\squared}* 300 = 1.8 \si{\milli\metre\squared} (\S\ref{ss:transactions})\\
  \hline
  Overhead for 5-block PIFO mesh & (5.55 + 1.8) / 200.0 = 3.7 \% \\
  \hline
  \end{tabular}
\end{small}
\caption{A 5-block PIFO mesh incurs a 3.7\% chip area overhead relative to
a baseline switch.}
\label{tab:area_overheads}
\end{table}

\paragraph{Varying parameters from the baseline design}

The flow scheduler has four parameters: rank width, metadata width, number of
logical PIFOs, and number of flows. Among these, increasing the number of flows
has the most impact on feasibility because the flow scheduler compares against
all flow entries in parallel. With other parameters set to their baseline
values, we vary the number of flows to determine the eventual limits of a flow
scheduler with today's transistor technology (Table~\ref{tab:num_flows}),
finding that we can scale up to 2048 flows while still meeting timing at 1 GHz.

\begin{table}
\begin{small}
\begin{tabular}{|p{0.1\textwidth}|p{0.1\textwidth}|p{0.2\textwidth}|}
\hline
\# of flows & Area (mm\textsuperscript{2}) & Meets timing at 1 GHz? \\
\hline
256 & 0.053 & Yes \\
\hline
512 & 0.107 & Yes \\
\hline
1024 & 0.224 & Yes \\
\hline
2048 & 0.454 & Yes \\
\hline
4096 & 0.914 & No \\
\hline
\end{tabular}
\end{small}
\caption{Flow scheduler area increases in proportion to the number of
flows and meets timing at 1 GHz until 2048 flows.}
\label{tab:num_flows}
\end{table}

The remaining parameters affect the area overhead of a flow scheduler, but have
little effect on whether or not the flow scheduler circuit meets timing. For
instance starting from the baseline design of the flow scheduler that takes up 0.224
\si{\milli\metre\squared} of area, increasing the rank width to 32 bits results
in an area of 0.317 \si{\milli\metre\squared}, increasing the number of logical
PIFOs to 1024 increases the area to 0.233 \si{\milli\metre\squared}, and
increasing the metadata width to 64 bits increases the area to 0.317
\si{\milli\metre\squared}. In all cases, the circuit continues to meet timing.

\subsection{Interconnecting PIFO blocks}
\label{ss:interconnect}

An interconnect between PIFO blocks is required for PIFO blocks to enqueue into
and dequeue from other PIFO blocks. Because the number of PIFO blocks is small,
we provide a full mesh between the PIFO blocks. Assuming a 5-block PIFO mesh,
this requires 5*4 = 20 sets of wires between PIFO blocks. Each set of
wires would need to carry all the inputs required for specifying an enqueue and
dequeue operation on a PIFO block.

We calculate the number of wires in each set for our baseline
design. For an enqueue operation, we require a logical PIFO ID (8
bits), the element's rank (16 bits), the element meta data (32 bits),
and the flow ID (10 bits). For a dequeue operation, we need a logical
PIFO ID (8 bits) and wires to store the dequeued element (32 bits).
This totals up to 106 bits per set of wires, or ~2120 bits across the
entire mesh. This is a relatively small number of wires and can easily
be supported on a chip. For example, RMT's match-action pipeline uses
nearly 2$\times$ the number of wires between each pair of pipeline
stages to move the packet header vector from one stage to the
next~\cite{rmt}. 

\section{Discussion}
\label{s:discussion}

\subsection{Buffer management}
Our design focuses on programmable scheduling and doesn't concern itself with
how the switch's data buffers are allocated to flows within a PIFO.  Buffer
management can either use static thresholds for each flow or dynamic thresholds
based on active queue management algorithms such as RED~\cite{red} and the
occupancy of other ports~\cite{broadcom_dynamic}.  In a shared-memory switch,
buffer management is largely orthogonal to scheduling, and is implemented using
counters that track the occupancies of various flows and ports. Before a packet
is enqueued into the scheduler, if any of these counters exceeds a
static or dynamic threshold, the packet is dropped. A similar
design could be used to check thresholds before enqueueing into a PIFO block as
well.

\subsection{Priority Flow Control}

Priority Flow Control (PFC)~\cite{pfc} is a standard that allows a switch to
send a {\em pause} message to an upstream switch requesting it to cease
transmission of packets belonging to a particular set of flows. PFC can be
integrated into our hardware design for PIFOs by masking out certain flows in
the flow scheduler during the dequeue operation if they have been paused
because of a PFC pause message and unmasking them when a PFC {\em resume}
message is received.

\subsection{Multi-pipeline switches}

The highest end switches today, such as the Broadcom Tomahawk~\cite{tomahawk},
support aggregate capacities exceeding 3 Tbit/sec. At a minimum packet size
of 64 bytes, this corresponds to an aggregate forwarding requirement of 6
billion packets/s. Because a single switch pipeline typically runs at 1 GHz and
processes a billion packets/s, such switches require multiple ingress and
egress pipelines that share access to the scheduler subsystem alone.

In such multi-pipeline switches, each PIFO block needs to support multiple
enqueue and dequeue operations per clock cycle (as many as the number of
ingress and egress pipelines). This is because packets can be enqueued from any
of the input ports every clock cycle, and each input port could reside in any
of the ingress pipelines. Similarly, each egress pipeline requires a new packet
every clock cycle, resulting in multiple dequeues every clock cycle.

A full-fledged design for multi-pipeline switches is beyond this
paper, but our current design does facilitate a multi-pipeline
implementation. A rank store supporting multiple pipelines is similar
to what is required in the data buffer of multi-pipeline switches
today. Building a flow scheduler to support multiple enqueues/dequeues
per clock is relatively easy because it is maintained in flip flops,
where it is simple to add multiple ports (unlike SRAM).

\section{Related Work}

\textbf{The Push-in First-out Queue:}
PIFOs were first introduced by Chuang et al.~\cite{pifo} as a proof construct
to prove that a combined input-output queued switch could emulate an
output-queued switch running different scheduling algorithms. The same paper
also shows how WFQ and strict priorities are specific instantiations of a PIFO.
We demonstrate that PIFOs can be used as an abstraction for programming
scheduling algorithms beyond WFQ and strict priorities, that they can be
composed together to program hierarchical scheduling algorithms, and finally
that they are feasible in today's transistor technology.

\textbf{Packet scheduling algorithms:}
Many scheduling algorithms~\cite{drr, srpt, stfq, wfq, hpfq, lstf, stopngo,
pFabric} have been proposed in the literature. Yet, only a handful (DRR,
traffic shaping, and strict priorities) are found in routers.  Even
programmable switches~\cite{rmt, flexpipe, xpliant} treat packet scheduling as
a black box. As shown in \S\ref{s:expressive}, PIFOs allow us to program these
and other as-yet-unknown scheduling algorithms, without the power, area, and
performance penalties of prior proposals~\cite{nosilverbullet} that require
fully reconfigurable FPGAs.

\begin{figure}
  \centering
  \includegraphics[width=0.7\columnwidth]{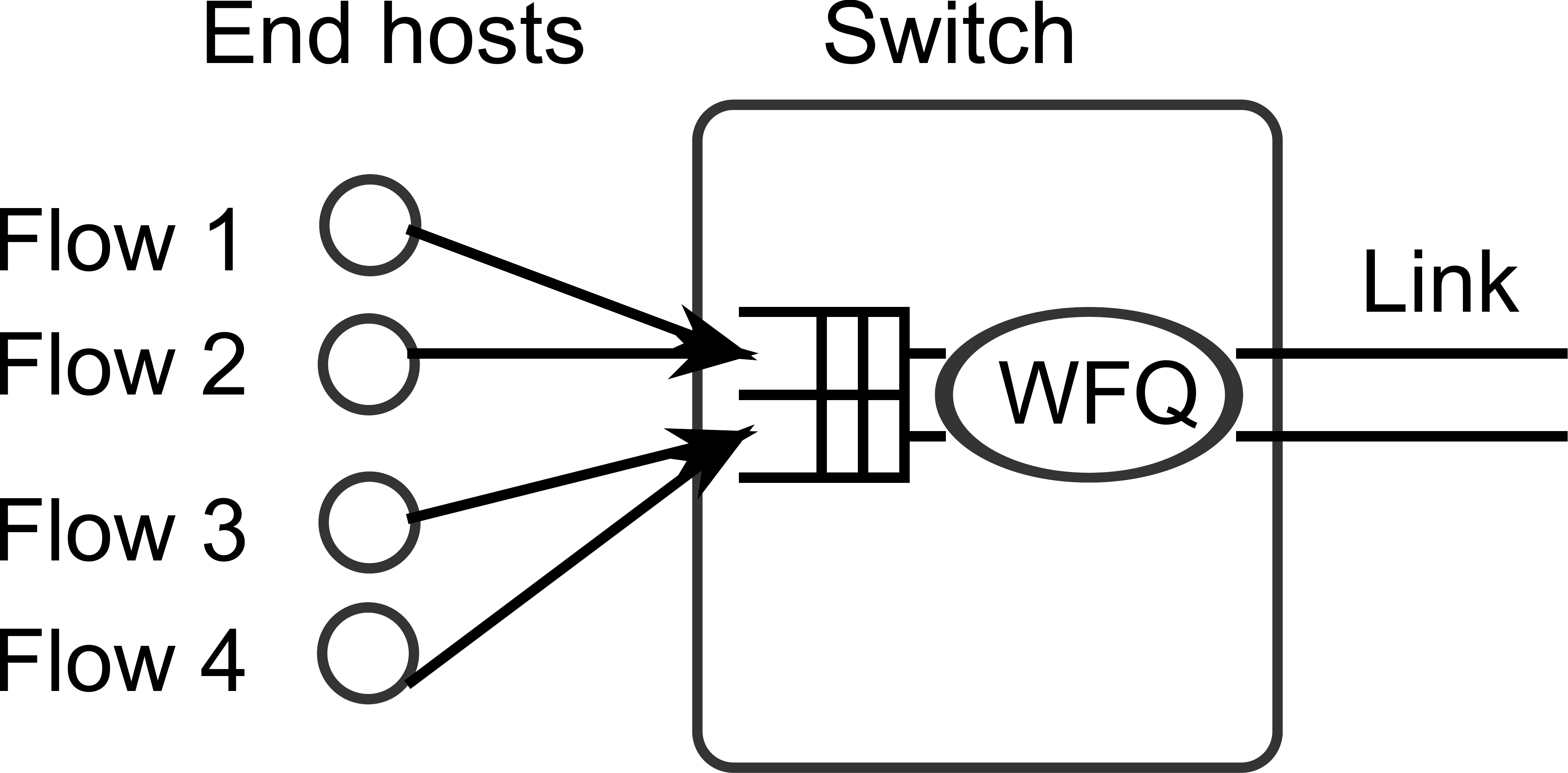}
  \caption{A switch's scheduling algorithm, such as WFQ, might aggregate flows
  from different end hosts into a single flow at the switch for the purpose of
  scheduling.}
  \label{fig:state}
\end{figure}

\textbf{Universal Packet Scheduling (UPS):} UPS~\cite{ups} uses the LSTF
algorithm by appropriately initializing slack times at end hosts and proves
that LSTF is universal under the strong assumption that packet departure times
for a scheduling algorithm are known up front.  LSTF is expressible using
PIFOs, but the set of schemes practically expressible with LSTF is itself
limited. For example, LSTF cannot express:
\begin{CompactEnumerate}
\item Hierarchical scheduling algorithms such as HPFQ, because it
  uses only one priority queue.
\item Non-work-conserving algorithms, because for such algorithms
  LSTF must know the departure time of each packet up-front, which
  is not practical.
\item Short-term bandwidth fairness in fair queueing, because LSTF
  maintains no switch state except one priority queue. As shown in
  Figure~\ref{fig:sched_trans}, programming a fair queueing
  algorithm requires us to maintain a virtual time state variable
  that is updated when packets are dequeued. Without this variable,
  a new flow could have arbitrary start times, and be deprived of
  its fair share indefinitely.
\item Scheduling policies that aggregate flows from distinct
  endpoints into a single flow at the switch
  (Figure~\ref{fig:state}), because LSTF provides no ability to
  maintain and update switch state progammatically.
\end{CompactEnumerate}

\textbf{Hardware designs for priority queues:} Hardware designs for a priority
queue have been proposed in the past~\cite{bhagwan, pheap}. These designs
typically employ a heap and scale to a large number of entries. They are the
basis for hierarchical schedulers in many deep-buffered core routers. However,
they occupy significant area---enough to warrant a dedicated chip for
the scheduler alone. They are unlikely to be feasible on merchant-silicon
shared-memory switching chips where chip area is at a premium. In contrast, our
design for the PIFO exploits two observations. First, there is considerable
structure in the arrival stream of ranks: ranks within a flow strictly increase
with time.  Second, the buffering requirements for shared-memory switches today
are much lesser than the buffering requirements of a core router. This permits
a simpler design relative to heaps.

\section{Conclusion}
\label{s:conclusion}

Until recently, it was widely assumed that the fastest switch chips would be
fixed-function; a programmable device could not have the same performance.
Recent research into programmable parsers~\cite{glen_parsing}, fast
programmable protocol-independent switch chips~\cite{rmt}, and languages to
program them~\cite{p4, pof}, coupled with a recent 3.2 Tbit/s programmable
commercial switch chip~\cite{xpliant} suggests that change might be afoot. But
so far, it has been considered off-limits to program the packet scheduler---in
part because the desired algorithms are so varied, and because the scheduler
sits right at the heart of the shared packet buffer where timing requirements
are tightest. It has been widely assumed too hard to find a useful abstraction
that can also be implemented in fast hardware.

PIFO appears to be a very promising abstraction: it includes a wide variety of
existing algorithms, and allows us to express new ones. We have found it
possible to implement PIFOs at line-rate with modest chip area overhead. We
believe the most exciting consequence will be the creation of many new
schedulers, invented by those who build and operate networks, iterated and
refined, then deployed for their own needs. No longer will research experiments
be limited to simulation and progress constrained by a vendor's choice of
scheduling algorithms. For those who need a new algorithm, they could create
their own, or might even download one from an open-source repository or a
reproducible SIGCOMM paper. To get there, we will need real switch chips with
PIFO schedulers we can program. The good news is that we see no reason why a
future generation of switch chips can not include a programmable PIFO
scheduler.

\balance
{\normalsize \bibliographystyle{acm}
\bibliography{paper}}

\begin{thebibliography}{10}

\bibitem{trident}
High {C}apacity {S}trata{XGS}\textregistered {T}rident {II} {E}thernet {S}witch
  {S}eries.
\newblock
  \url{http://www.broadcom.com/products/Switching/Data-Center/BCM56850-Series}.

\bibitem{tomahawk}
High-density 25/100 gigabit ethernet {StrataXGS} tomahawk ethernet switch
  series.
\newblock
  \url{http://www.broadcom.com/products/Switching/Data-Center/BCM56960-Series}.

\bibitem{int}
"in-band network telemetry".
\newblock \url{https://github.com/p4lang/p4factory/tree/master/apps/int}.

\bibitem{flexpipe}
Intel {F}lex{P}ipe.
\newblock
  \url{http://www.intel.com/content/dam/www/public/us/en/documents/product-briefs/ethernet-switch-fm6000-series-brief.pdf}.

\bibitem{pfc}
"priority flow control: Build reliable layer 2 infrastructure".
\newblock
  \url{http://www.cisco.com/en/US/prod/collateral/switches/ps9441/ps9670/white_paper_c11-542809_ns783_Networking_Solutions_White_Paper.html}.

\bibitem{sram_estimate}
Sram - arm.
\newblock
  \url{https://www.arm.com/products/physical-ip/embedded-memory-ip/sram.php}.

\bibitem{tbf}
Token {B}ucket.
\newblock \url{https://en.wikipedia.org/wiki/Token_bucket}.

\bibitem{xpliant}
{XP}liant\texttrademark {E}thernet {S}witch {P}roduct {F}amily.
\newblock
  \url{http://www.cavium.com/XPliant-Ethernet-Switch-Product-Family.html}.

\bibitem{pFabric}
{\sc Alizadeh, M., Yang, S., Sharif, M., Katti, S., McKeown, N., Prabhakar, B.,
  and Shenker, S.}
\newblock p{F}abric: {M}inimal {N}ear-optimal {D}atacenter {T}ransport.
\newblock In {\em SIGCOMM\/} (2013).

\bibitem{hpfq}
{\sc Bennett, J. C.~R., and Zhang, H.}
\newblock Hierarchical {P}acket {F}air {Q}ueueing {A}lgorithms.
\newblock In {\em SIGCOMM\/} (1996).

\bibitem{bhagwan}
{\sc Bhagwan, R., and Lin, B.}
\newblock Fast and {S}calable {P}riority {Q}ueue {A}rchitecture for
  {H}igh-{S}peed {N}etwork {S}witches.
\newblock In {\em In Proceedings of Infocom 2000\/} (March 2000), IEEE
  Communications Society.

\bibitem{p4}
{\sc Bosshart, P., Daly, D., Gibb, G., Izzard, M., McKeown, N., Rexford, J.,
  Schlesinger, C., Talayco, D., Vahdat, A., Varghese, G., and Walker, D.}
\newblock {P4}: {P}rogramming {P}rotocol-independent {P}acket {P}rocessors.
\newblock {\em SIGCOMM Comput. Commun. Rev. 44}, 3 (July 2014), 87--95.

\bibitem{rmt}
{\sc Bosshart, P., Gibb, G., Kim, H.-S., Varghese, G., McKeown, N., Izzard, M.,
  Mujica, F., and Horowitz, M.}
\newblock Forwarding {M}etamorphosis: {F}ast {P}rogrammable {M}atch-action
  {P}rocessing in {H}ardware for {SDN}.
\newblock In {\em SIGCOMM\/} (2013).

\bibitem{broadcom_dynamic}
{\sc Choudhury, A.~K., and Hahne, E.~L.}
\newblock Dynamic queue length thresholds for shared-memory packet switches.
\newblock {\em IEEE/ACM Trans. Netw. 6}, 2 (Apr. 1998), 130--140.

\bibitem{pifo}
{\sc Chuang, S.-T., Goel, A., McKeown, N., and Prabhakar, B.}
\newblock Matching output queueing with a combined input/output-queued switch.
\newblock {\em Selected Areas in Communications, IEEE Journal on 17}, 6 (Jun
  1999), 1030--1039.

\bibitem{intserv}
{\sc Clark, D.~D., Shenker, S., and Zhang, L.}
\newblock Supporting real-time applications in an integrated services packet
  network: Architecture and mechanism.
\newblock In {\em Conference Proceedings on Communications Architectures \&Amp;
  Protocols\/} (New York, NY, USA, 1992), SIGCOMM '92, ACM, pp.~14--26.

\bibitem{wfq}
{\sc Demers, A., Keshav, S., and Shenker, S.}
\newblock {A}nalysis and {S}imulation of a {F}air {Q}ueueing {A}lgorithm.
\newblock In {\em SIGCOMM\/} (1989).

\bibitem{red}
{\sc Floyd, S., and Jacobson, V.}
\newblock Random early detection gateways for congestion avoidance.
\newblock {\em IEEE/ACM Trans. Netw. 1}, 4 (Aug. 1993), 397--413.

\bibitem{cbq}
{\sc Floyd, S., and Jacobson, V.}
\newblock Link-sharing and resource management models for packet networks.
\newblock {\em IEEE/ACM Trans. Netw. 3}, 4 (Aug. 1995), 365--386.

\bibitem{cbq_impl}
{\sc Floyd, S., and Speer, M.~F.}
\newblock Experimental results for class-based queueing.

\bibitem{glen_parsing}
{\sc Gibb, G., Varghese, G., Horowitz, M., and McKeown, N.}
\newblock Design principles for packet parsers.
\newblock In {\em Architectures for Networking and Communications Systems
  (ANCS), 2013 ACM/IEEE Symposium on\/} (Oct 2013), pp.~13--24.

\bibitem{stopngo}
{\sc Golestani, S.~J.}
\newblock A stop-and-go queueing framework for congestion management.
\newblock In {\em Proceedings of the ACM Symposium on Communications
  Architectures \&Amp; Protocols\/} (New York, NY, USA, 1990), SIGCOMM '90,
  ACM, pp.~8--18.

\bibitem{stfq}
{\sc Goyal, P., Vin, H.~M., and Chen, H.}
\newblock Start-time {F}air {Q}ueueing: {A} {S}cheduling {A}lgorithm for
  {I}ntegrated {S}ervices {P}acket {S}witching {N}etworks.
\newblock In {\em SIGCOMM\/} (1996).

\bibitem{pheap}
{\sc Ioannou, A., and Katevenis, M.}
\newblock Pipelined {H}eap ({P}riority {Q}ueue) {M}anagement for {A}dvanced
  {S}cheduling in {H}igh-{S}peed {N}etworks.
\newblock {\em Networking, IEEE/ACM Transactions on 15}, 2 (April 2007),
  450--461.

\bibitem{tpp}
{\sc Jeyakumar, V., Alizadeh, M., Geng, Y., Kim, C., and Mazi\`{e}res, D.}
\newblock {M}illions of {L}ittle {M}inions: {U}sing {P}ackets for {L}ow
  {L}atency {N}etwork {P}rogramming and {V}isibility.
\newblock In {\em SIGCOMM\/} (2014).

\bibitem{eyeq}
{\sc Jeyakumar, V., Alizadeh, M., Mazi{\`e}res, D., Prabhakar, B., Greenberg,
  A., and Kim, C.}
\newblock Eye{Q}: {P}ractical {N}etwork {P}erformance {I}solation at the
  {E}dge.
\newblock In {\em Presented as part of the 10th USENIX Symposium on Networked
  Systems Design and Implementation (NSDI 13)\/} (Lombard, IL, 2013), USENIX,
  pp.~297--311.

\bibitem{hrr}
{\sc Kalmanek, C.~R., Kanakia, H., and Keshav, S.}
\newblock Rate controlled servers for very high-speed networks.
\newblock In {\em Global Telecommunications Conference, 1990, and
  Exhibition.'Communications: Connecting the Future', GLOBECOM'90., IEEE\/}
  (1990), IEEE, pp.~12--20.

\bibitem{lstf}
{\sc Leung, J.-T.}
\newblock A {N}ew {A}lgorithm for {S}cheduling {P}eriodic, {R}eal-{T}ime
  {T}asks.
\newblock {\em Algorithmica 4}, 1-4 (1989), 209--219.

\bibitem{sfq}
{\sc McKenney, P.}
\newblock Stochastic fairness queuing.
\newblock In {\em INFOCOM '90, Ninth Annual Joint Conference of the IEEE
  Computer and Communication Societies. The Multiple Facets of Integration.
  Proceedings, IEEE\/} (Jun 1990), pp.~733--740 vol.2.

\bibitem{ups}
{\sc Mittal, R., Agrawal, R., Ratnasamy, S., and Shenker, S.}
\newblock Universal {P}acket {S}cheduling.
\newblock In {\em Proceedings of the Fourteenth ACM Workshop on Hot Topics in
  Networks\/} (2015).

\bibitem{faircloud}
{\sc Popa, L., Kumar, G., Chowdhury, M., Krishnamurthy, A., Ratnasamy, S., and
  Stoica, I.}
\newblock Fair{C}loud: {S}haring the {N}etwork in {C}loud {C}omputing.
\newblock In {\em SIGCOMM\/} (2012).

\bibitem{sced}
{\sc Sariowan, H., Cruz, R.~L., and Polyzos, G.~C.}
\newblock {SCED}: {A} {G}eneralized {S}cheduling {P}olicy for {G}uaranteeing
  {Q}uality-of-service.
\newblock {\em IEEE/ACM Trans. Netw. 7}, 5 (Oct. 1999), 669--684.

\bibitem{srpt}
{\sc Schrage, L.~E., and Miller, L.~W.}
\newblock {T}he {Q}ueue {M/G/1} with the {S}hortest {R}emaining {P}rocessing
  {T}ime {D}iscipline.
\newblock {\em Operations Research 14}, 4 (1966), 670--684.

\bibitem{drr}
{\sc Shreedhar, M., and Varghese, G.}
\newblock Efficient {F}air {Q}ueuing using {D}eficit {R}ound {R}obin.
\newblock {\em Networking, IEEE/ACM Transactions on 4}, 3 (1996), 375--385.

\bibitem{domino_arxiv}
{\sc Sivaraman, A., Budiu, M., Cheung, A., Kim, C., Licking, S., Varghese, G.,
  Balakrishnan, H., Alizadeh, M., and McKeown, N.}
\newblock Packet transactions: {A} programming model for data-plane algorithms
  at hardware speed.
\newblock {\em CoRR abs/1512.05023\/} (2015).

\bibitem{pifo_hotnets}
{\sc Sivaraman, A., Subramanian, S., Agrawal, A., Chole, S., Chuang, S.-T.,
  Edsall, T., Alizadeh, M., Katti, S., McKeown, N., and Balakrishnan, H.}
\newblock Towards programmable packet scheduling.
\newblock In {\em Proceedings of the 14th ACM Workshop on Hot Topics in
  Networks\/} (New York, NY, USA, 2015), HotNets-XIV, ACM, pp.~23:1--23:7.

\bibitem{nosilverbullet}
{\sc Sivaraman, A., Winstein, K., Subramanian, S., and Balakrishnan, H.}
\newblock No {S}ilver {B}ullet: {E}xtending {SDN} to the {D}ata {P}lane.
\newblock In {\em Proceedings of the Twelfth ACM Workshop on Hot Topics in
  Networks\/} (2013).

\bibitem{pof}
{\sc Song, H.}
\newblock Protocol-oblivious {F}orwarding: {U}nleash the {P}ower of {SDN}
  {T}hrough a {F}uture-proof {F}orwarding {P}lane.
\newblock In {\em HotSDN\/} (2013).

\bibitem{jitteredd}
{\sc Verma, D., Zhang, H., and Ferrari, D.}
\newblock Guaranteeing delay jitter bounds in packet switching networks.
\newblock In {\em Proceedings of TRICOMM\/} (1991), vol.~91, pp.~35--46.

\bibitem{rcsd}
{\sc Zhang, H., and Ferrari, D.}
\newblock {Rate-Controlled Service Disciplines}.
\newblock {\em J. High Speed Networks 3}, 4 (1994), 389--412.

\end{thebibliography}

\end{document}